\documentclass{egpubl}

\usepackage[T1]{fontenc}
\usepackage{dfadobe}  

\usepackage{cite}  
\BibtexOrBiblatex

\electronicVersion
\PrintedOrElectronic
\ifpdf \usepackage[pdftex]{graphicx} \pdfcompresslevel=9
\else \usepackage[dvips]{graphicx} \fi

\usepackage{egweblnk}

\usepackage[dvipsname,svgnames]{xcolor}
\usepackage{adjustbox}
\usepackage{tikz}

\usepackage{balance}

\usepackage[caption=false,labelfont=sf,textfont=sf]{subfig}

\usepackage{amsmath,amsfonts}
\usepackage{array}
\usepackage{enumitem}
\usepackage{textcomp}
\usepackage{stfloats}
\usepackage{url}
\usepackage{verbatim}
\usepackage{graphicx}
\usepackage{cite}

\usepackage{tabu}                      
\usepackage{booktabs}                  
\usepackage{lipsum}                    
\usepackage{mwe}                       
\usepackage[capitalise,noabbrev]{cleveref}

\usepackage{algorithm}
\usepackage{algpseudocode}
\usepackage{algorithmicx}
\makeatletter
\renewcommand*{\ALG@name}{Pipeline}   
\makeatother

\usepackage{makecell}                 
\usepackage{xcolor}                   

\usepackage{float}

\title[Feature-Action Design Patterns for Storytelling]%
        {Feature-Action Design Patterns \\ for Storytelling Visualizations with Time Series Data}

\author[Khan et al.]        
{\parbox{\textwidth}{\centering 
S. Khan$^{1}$\orcid{0000-0002-6796-5670}, 
S. Jones$^2$\orcid{0000-0002-2191-3170}, 
B. Bach$^3$\orcid{0000-0002-9201-7744},
J. Cha$^2$\orcid{0000-0002-2498-4214},
M. Chen$^2$\orcid{0000-0001-5320-5729},
J. Meikle$^2$,
J. C. Roberts$^4$\orcid{0000-0001-7718-3181},\\
J. Thiyagalingam$^2$\orcid{0000-0002-2167-1343},
J. Wood$^5$\orcid{0000-0001-9270-247X},
P. D. Ritsos$^4$\orcid{0000-0001-9308-3885}
        \\
{\parbox{\textwidth}{\centering 
        $^1$Science and Technology Facilities Council (STFC), UK, $^2$University of Oxford, UK\\
        $^3$University of Edinburgh, UK, $^4$Bangor University, UK, $^5$City University, UK} 
        }}}

\definecolor{featureViolet}{RGB}{128, 128,255}
\definecolor{actionGreen}{RGB}{0,190,191}

\definecolor{workflowOrange}{RGB}{253, 234, 218}
\definecolor{workflowOlive}{RGB}{215, 228, 189}
\definecolor{workflowBlue}{RGB}{220,230,242}
\definecolor{workflowGray}{RGB}{242,242,242}
\definecolor{workflowPurple}{RGB}{179,162,199}
\definecolor{workflowRed}{RGB}{222,164,162}

\newcommand{\covid}{COVID-19}

\DeclareRobustCommand\myRounded[1]{\tikz{\draw[fill=#1, rounded corners=1pt]  rectangle (.9em,.6em);}}
\DeclareRobustCommand\mySquared[1]{\tikz{\draw[fill=#1]  rectangle (.6em,.6em);}}

\DeclareRobustCommand\featureP{\tikz[baseline=-0.9ex]{\draw[fill=featureViolet, featureViolet, line width=0.0pt] circle (0.125cm) node[text=white] {\tiny\textsf{F}};}}
 
\DeclareRobustCommand\actionP{\tikz[baseline=-0.9ex]{\draw[fill=actionGreen, actionGreen, line width=0.0pt] circle (0.125cm) node[text=white] {\tiny\textsf{A}};}}

\DeclareRobustCommand\feature{\raisebox{-.1em}{\maxsizebox{1em}{2ex}{\featureP}}}
\DeclareRobustCommand\action{\raisebox{-.1em}{\maxsizebox{1em}{2ex}{\actionP}}}

\newcommand{\rev}[1]{{#1}} 
\newcommand{\old}[1]{{}} 

\begin{document}

\maketitle
\begin{abstract}
We present a method to create storytelling visualization with time series data. Many \textit{personal} decisions nowadays rely on access to dynamic data regularly, as we have seen during the \covid{} pandemic. It is thus desirable to construct storytelling visualization for dynamic data that is selected by an individual for a specific context. Because of the need to tell data-dependent stories, predefined storyboards based on known data cannot accommodate dynamic data easily nor scale up to many different individuals and contexts.
Motivated initially by the need to communicate time series data during the \covid{} pandemic, we developed a novel computer-assisted method for meta-authoring of stories, which enables the design of storyboards that include feature-action patterns in anticipation of potential features that may appear in dynamically arrived or selected data.
In addition to meta-storyboards involving \covid{} data, we also present  storyboards for telling stories about progress in a machine learning workflow.    
Our approach is complementary to traditional methods for authoring storytelling visualization, and provides an efficient means to construct data-dependent storyboards for different data-streams of similar contexts.


\end{abstract}  

\section{Introduction}
\label{SEC:intro}

\begin{figure}[t]
  \centering
  \includegraphics[width=\columnwidth]{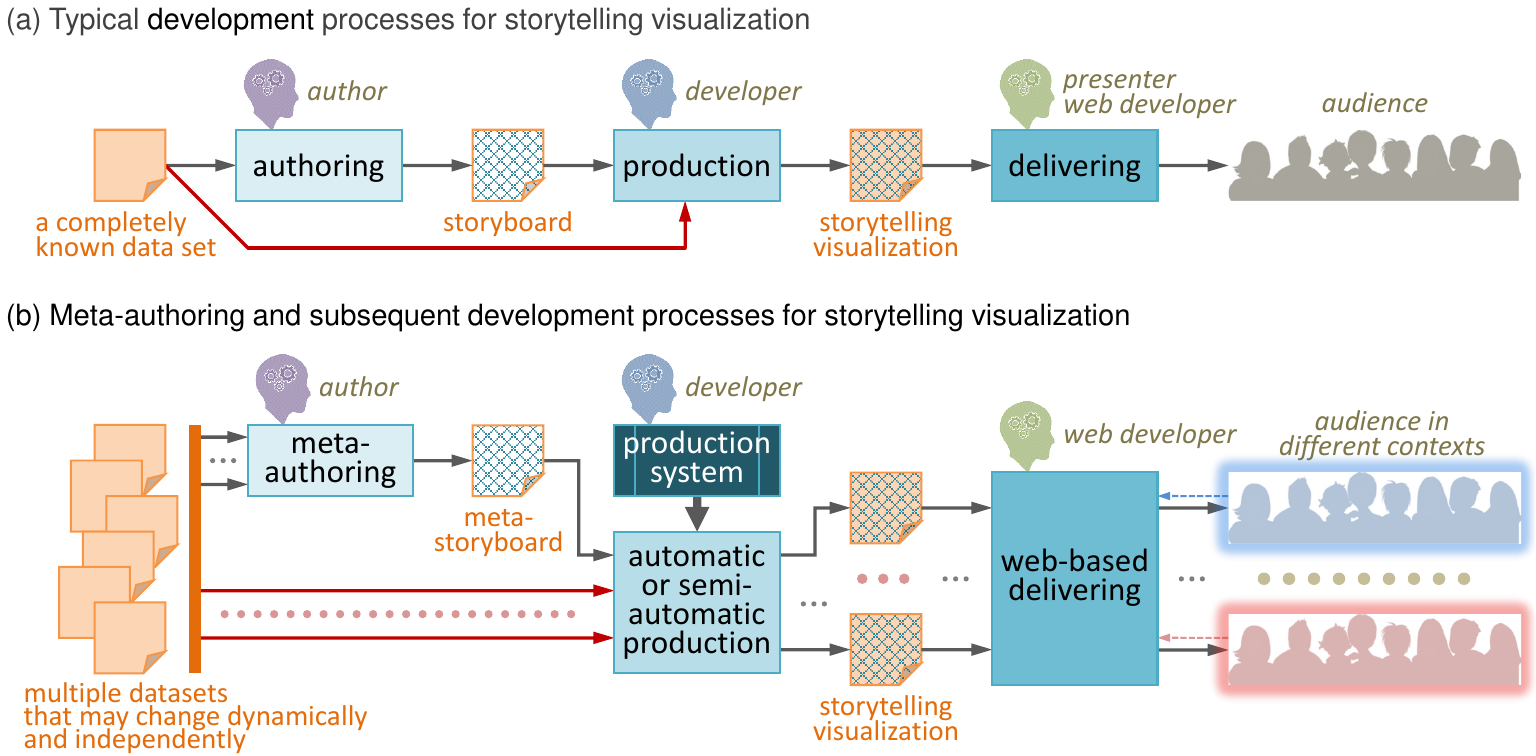}
  \vspace{-3mm}
  \caption{\rev{(a) In a typical workflow for creating storytelling visualization, an author defines a storyboard for a known dataset, which is then developed as a web-based visualization, usually for a specific target audience. (b) With our approach, the author creates a meta-storyboard that works with multiple, dynamic, and often not-yet-inspected datasets. The storyboard is converted by a developer, following rules that facilitate the automatic or semi-automatic depiction of user-driven stories, for different target audiences.}\vspace{-2mm}}
  \label{FIG:overview}
  \vspace{-4mm}
\end{figure}

Visualization provides a powerful means for telling stories about data \cite{KosaraMackinlay2013,GershonWard2001, Tong2018StorytellingAV, BongshinETAL2015}. In most workflows, the authors and the developers of storytelling visualization are typically given a complete dataset that is not expected to change after the storyboard is constructed (e.g.,~\cite{shi_calliope_2021,Offenwanger:2024:TVCG}). \cref{FIG:overview}(a) illustrates such a workflow. With the rapid growth of data volume and context, it is increasingly important to deliver data-driven storytelling visualization to different audiences. Moreover, individuals will likely pay more attention to stories relevant to them, such as those about happenings close to their location, information affecting their decisions, or historical facts interesting to them.  

Such requirements became strikingly noticeable during the \covid{} pandemic. Institutions, such as governments and news outlets, were able to transfer data at the national level to stories effectively. Nevertheless, many members of the public found it difficult to access information of more immediate interest, such as finding out what happened in their location, checking recent data related their planned journeys, or comparing historical data between regions of interest. It is not scalable to author many different storyboards for every individual region or every pair of regions. Nor is it scalable to change storyboards manually, whenever a dynamic dataset is updated with newly-arrived data. Moreover, having multiple storyboards and changing them frequently can be tedious for developers who implement storytelling visualization software.

As illustrated in \cref{FIG:overview}(b), ideally, authors of storytelling visualization could construct a common storyboard, for multiple datasets that may change dynamically and independently. When the common storyboard is applied to multiple datasets at a particular moment (e.g., COVID-19 data streams for different regions in a given period), there could be an efficient mechanism to generate different storytelling visualization for audiences in different contexts (e.g., in different regions). We refer to such a common storyboard as a \emph{meta-storyboard}, while we refer to the process of creating it as \emph{meta-authoring} (outlined in~\cref{FIG:metaprocess}).

There are many technical challenges in enabling the development workflow in \cref{FIG:overview}(b), including devising software tools to support meta-authoring, mechanisms for mapping meta-storyboards to different datasets to generate different stories, and software platforms to deliver the resulting storytelling visualizations to audiences. 
Example outputs are frequently updated storytelling visualizations for different regions and/or periods, both selected by the users. 
In such cases, if a developer had to read a storyboard and manually transform it to a storytelling visualization for each data stream and whenever new data arrives, it would not be efficient or even feasible as the same process had to be repeated again and again. 
It is thus desirable to have an automatic or semi-automatic mechanism to combine a meta-storyboard with datasets that may be changing. 

While such difficulties may be caused by the differences between datasets and time frames, many data patterns in these datasets are expected to be similar. A meta-storyboard is defined for a group of notionally-similar datasets, such as the time series of \covid{} daily cases in different regions. Moreover, similar data patterns can be depicted using similar visual patterns. The key to developing an efficient mechanism for combining a meta-storyboard with individual datasets is to define, implement, and apply design patterns for storytelling visualization~\cite{bach2018narrative}. To avoid overusing the word ``pattern'', we use the term ``data feature'' in place of ``data pattern'', and ``visualization action'' in place of ``visual pattern''. Hence the design patterns in storytelling visualization are patterns of mapping feature to action, i.e., ``feature-action'' design patterns.

As part of the RAMPVIS project ~\cite{Chen:2022:E,Dykes:2022:PTRSA} for providing a variety of visualization capabilities to support epidemiological modelling, a small team focused on devising novel techniques for storytelling visualization, which enables meta-authoring, by providing a production workflow as shown in \cref{FIG:overview}(b). Our method is a novel addition to the emerging set of methods for automated storytelling (e.g.,~\cite{wangDatashot2020, Roslingifier2022}), and offers the first scalable solution to the meta-authoring of stories about multiple dynamic time series datasets.
While our approach was developed in the context of the \covid{} pandemic, it is generalisable beyond this context, e.g., for visualising stories on carbon emissions, plastic waste, personal financial spending, and so on. To demonstrate the feasibility, we repurposed the software to provide storytelling visualization of machine learning (ML) workflows.

Therefore, our contributions are: (i) meta-authoring as a new method for creating storyboards for dynamic multi-stream data, (ii) an algorithmic pipeline for using pre-defined feature-action patterns for realising meta-storyboards in response to unseen data (accompanied by supplementary material), and (iii) six storyboards of two case studies, as demonstration.

\section{Related Work}
\label{SEC:background}
Storytelling and visualization have a long history~\cite{Tong2018StorytellingAV}, helping to reveal information in ways that are intuitive and compelling~\cite{GershonWard2001}. Employed techniques are wide ranging from sketching~\cite{Lee2013SketchStory}, slide shows~\cite{hullman2013deeper}, comics~\cite{wang2021interactive} to investigating specific tasks in storytelling, such as linking~\cite{zhi2019linking}, collaboration~\cite{BongshinETAL2015}, immersion~\cite{Williams-et-al-VIS-2020} and learning methods~\cite{TangETAL2021}.

\subsection{Personalising Information Visualization}

As many of our daily activities are mediated by some form of interactive technology, recording, sharing and utilising data and related information about said activities has become more prevalent. In particular, data-driven information that can be personal, i.e., concern us either as individuals or some identifiable collective, is already driving our decision making. For instance, Yousuf and Conlan explore personal visual-learning narratives~\cite{YousufConlan2018}. During the COVID-19 pandemic, the degree of the technological mediation increased, either as we started working and socialising more online, or as we tried to follow the pandemic's progression through visualizations on public media. Looking at daily, national or local infection rates, became a daily activity which often determined our daily or monthly routine and choices. 

Nevertheless, most of the visualizations that we had access to provided either none or limited functionality, when it came to enabling personalised points of view. For instance, dashboards such as the John Hopkins \covid{} map~\cite{JohnHopkins} provided only a world and national overview, whereas Governmental portals, such as the UK's Coronavirus dashboard~\cite{GovUK} allowed the selection of locations (via post code) but did not provide any form of contextualisation, or a story. This is also true for visualization interfaces that presents data on world events, which nonetheless affect our day to day lives (e.g., world crises and impact on food pricing) and often remain on national level averages. 

Inevitably, providing more personalised storytelling requires: a) the availability of data for a more personal point of view, e.g., specific to a location, person, communities etc. and b) mechanisms to involve the users, and allow them to tailor their queries, explore the resulting visual depictions, and involve and engage them on a suitably personal level. Our work takes steps towards this direction.

\subsection{Storytelling Concepts (especially Temporal Data)}

Data-driven storytelling aims to provide a curated lens on evidence in data. It does so through contextual information, a curated set of messages, compelling narrative devices~\cite{satyanarayan2014authoring}, and often a sequential ordering of information. It employs narrative patterns~\cite{bach2018narrative}, such as \textit{gradual reveal}, or \textit{juxtaposition}, and communicates through a set of genres~\cite{segel2010narrative} such as videos~\cite{amini2016authoring}, Data GIFs~\cite{shu2020makes}, slideshows~\cite{hullman2013deeper,QianwenETAL2019}, or data comics~\cite{bach_design_2018}. Visual narratives help to explain ideas~\cite{RobertsETAL2022}, which are especially useful in education settings~\cite{YousufConlan2018}. 
While studying data comics, Wang et al.~\cite{wang2019comparing} found that by breaking down the complexity of information into individual steps, and presenting them sequentially (i.e., comics), readers could follow the story better, compared to infographics or illustrated texts. Feature-action data patterns create such sequences which a reader can navigate forwards and backwards.

A key point in storytelling is the relation between the story and the audience. In author-driven stories, narration, sequence, and content are defined by the author of the story, with little to no agency from the audience. In reader-driven storytelling, readers gain agency over the story and consequently can personalise it. Agency usually is achieved through interaction, e.g., by navigating within the story, data selection or free exploration~\cite{wang2021interactive}. However, one of the drawbacks of interactive storytelling is the discoverability of interaction affordances~\cite{cox2011editing,boy2015suggested}, risking to impede personalised storytelling~\cite{mckenna2017visual}. With our approach, we reduce the need for interaction and customization by creating prescribed stories for a given set of input data. As shown in~\cref{SEC:usecases}, we keep interaction to a minimum and instead provide stories at different levels of granularity. While the sequence, narration, and contents is fixed, said granularity, which is informed by our algorithm (\cref{SEC:Algorithm}), defines how many messages the story has.

Storytelling of temporal data has a long tradition as demonstrated through two comprehensive surveys~\cite{rosenberg2013cartographies,BrehmerETAL2017ExpressiveStorytelling} which feature many historic examples of time-\textit{lines}. Through their sequential nature, temporal data lends itself particularly well to sequential and author-driven storytelling~\cite{WalkerETALStoryboarding2015}.
However, these bespoke examples make generalisation to other datasets hard. 
Commercial tools for creating visual timelines include TimelineJS~\cite{timelineJS} and most recently
Timeline Storyteller~\cite{BrehmerETAL2017ExpressiveStorytelling}.
Likewise, Timeline Curator supports the creation of visual timelines from text, extracting data from the latter, and enabling the curation of the visual timeline~\cite{fulda2015timelinecurator}. 

\subsection{Automation in Storytelling}

A good number of systems allow authors to create storytelling visualizations interactively, e.g., \cite{Offenwanger:2024:TVCG,Moerth:2024:TVCG}.
Meanwhile automation for storytelling aims at increasing the access to information by lowering the burden for humans to create bespoke stories. 
Some systems aim more at supporting analysts gaining an overview over their data and personalising their interest through interaction~\cite{shi_calliope_2021}, whereas other systems aim to support the creation of effective and compelling \textit{communication} material~\cite{wangDatashot2020}.
Other systems deal with fact extraction, as by Law et al.~\cite{Law2020_insights,Shen:2024:TVCG}. Facts can be selected based on statistics and ordered based on a ranking score representing subjective relevance, interesting-ness, and importance~\cite{shi_calliope_2021,wangDatashot2020}.
User interaction can enable dynamic data selection \cite{Wu:2024:TVCG}. 
Our method works in a similar manner in that it provides a specific routine for determining specific features, while ranking is defined by a Gaussian function. To the best of our knowledge, all these systems explicitly focus on tabular data. In this paper, we consider storytelling visualization of temporal data. 

Storytelling systems may group visual information into panels based on category~\cite{Demiralp2017_foresight},
create dashboards~\cite{wu2021multivision}, or generate infographic-like fact sheets of visualizations around topics, featuring textual explanations~\cite{wangDatashot2020}. To organise data facts into sequences, graph-based approaches create a similarity graph from all facts and consequently select a sequence of visualizations (a path through the graph) based on minimising edge weight~\cite{yu2010automatic,kim_graphscape_2017}. Data facts can also be ordered using a logic-oriented Monte Carlo tree search algorithm~\cite{shi_calliope_2021}.

Closest to us, Parry et al. \cite{Parry2017_Snooker} presented an automated method to extract important frames from a video, for composing a sports game summary storyboard. They used a Gaussian mixture model to estimate the importance of individual frames. We adapted their technique for time series data. 
With our approach, data facts are extracted from the time series through analysis as well as complemented through public information on policy decisions. A Gaussian distribution defines which facts are shown for each given timeline and user-defined granularity. Consequently, timelines are build up step-by-step, progressing through identified breakpoints, while contextual information is displayed in textual labels. For our data, the narrative sequence follows the temporal order of breakpoints.

\section{Meta-authoring}
\label{SEC:usecases}

\begin{figure}[t]
  \centering
  \includegraphics[width=0.98\linewidth]{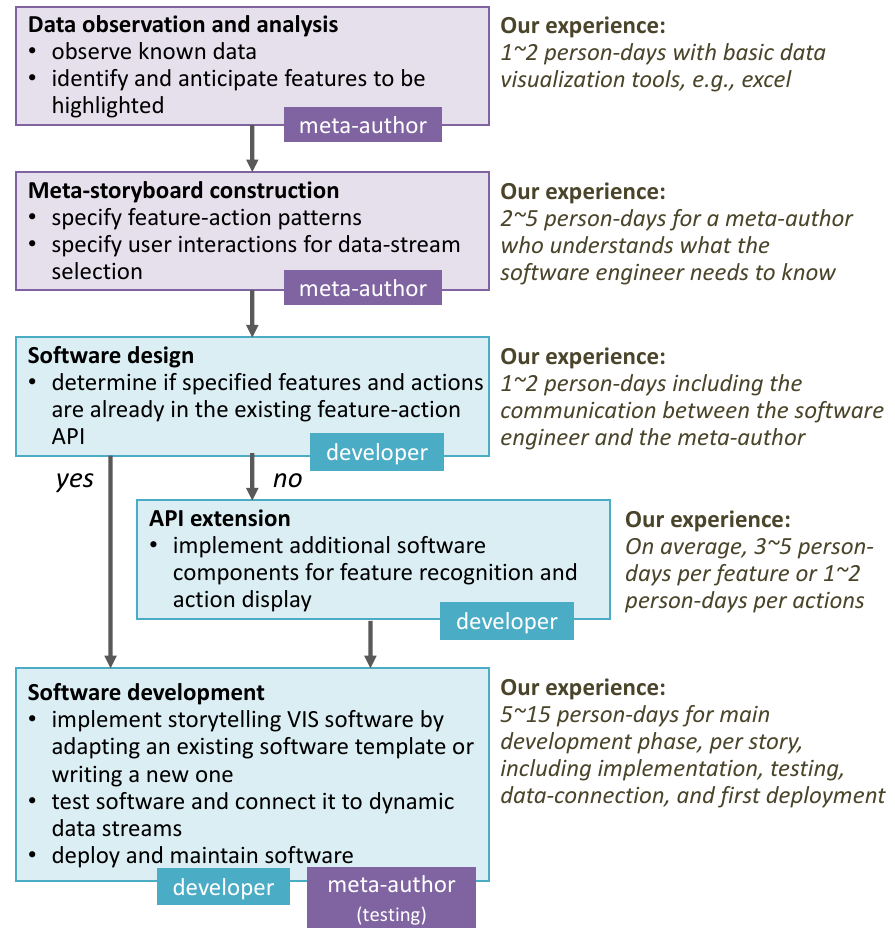}\vspace{-1mm}
  \caption{Overview of the proposed meta-authoring process, involving a story meta-author and a developer. The resulting storytelling visualization software can be applied to many similar, dynamic, and often not-yet-inspected data streams.}
  \label{FIG:metaprocess}
  \vspace{-4mm}
\end{figure}

Our meta-authoring process enables story authors to define narratives that can be applied in many similar, dynamic, and often not-yet-inspected datas treams. As outlined in \cref{FIG:metaprocess}, a meta-author needs to (1) explore the data; (2) turn specific story items into generalised data features~\feature{}, and (3) map features to actions~\action{}. A developer then uses feature-action APIs to transform a meta-storyboard into a piece of generalised visualization software. 

Firstly, a meta-author inspects some known data and ascertains key \textbf{features}~\feature{} anticipated in the data. One important challenge that meta-authors face is to cope with the quantity, range of, and precise values of the available data streams. Unlike in designing traditional storytelling visualization -- where authors and developers would have immediate access to the data -- in meta-authoring, a meta-author may not have (or do not need to have) seen all data in detail. Hence, it is critical for the meta-author to be able to anticipate the potential features that the data may or may not have, and create a story specification that can work either way. 

Subsequently, the storyteller needs to express the \textbf{features}~\feature{} in a descriptive and generalisable form, that allows the detection of said features without requiring knowledge of when (or whether) they occur in each data stream. An example would be ``\textit{find if the data in each month featuring a rising slope above 15$^\circ$}''.

\begin{figure}[t]
\centering
    \vspace{-2mm}
    \subfloat[highlight]{\includegraphics[width=1.5cm]{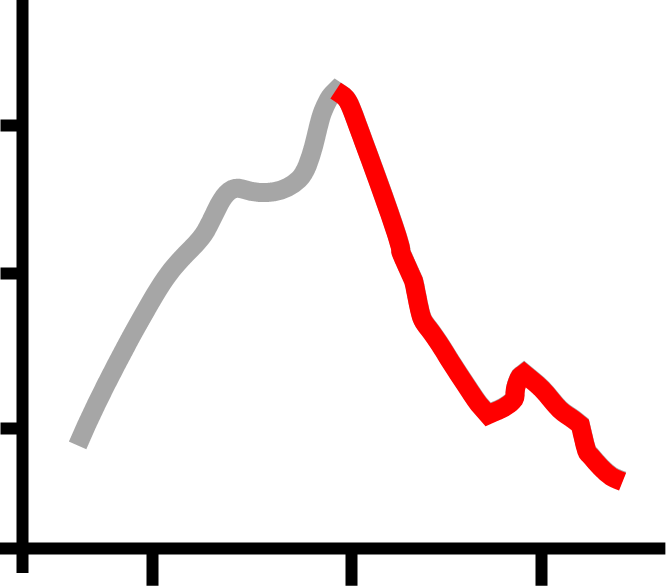}\label{SUBFIG:action1}}
      \hspace{3mm}
     \subfloat[label]{\includegraphics[width=2cm]{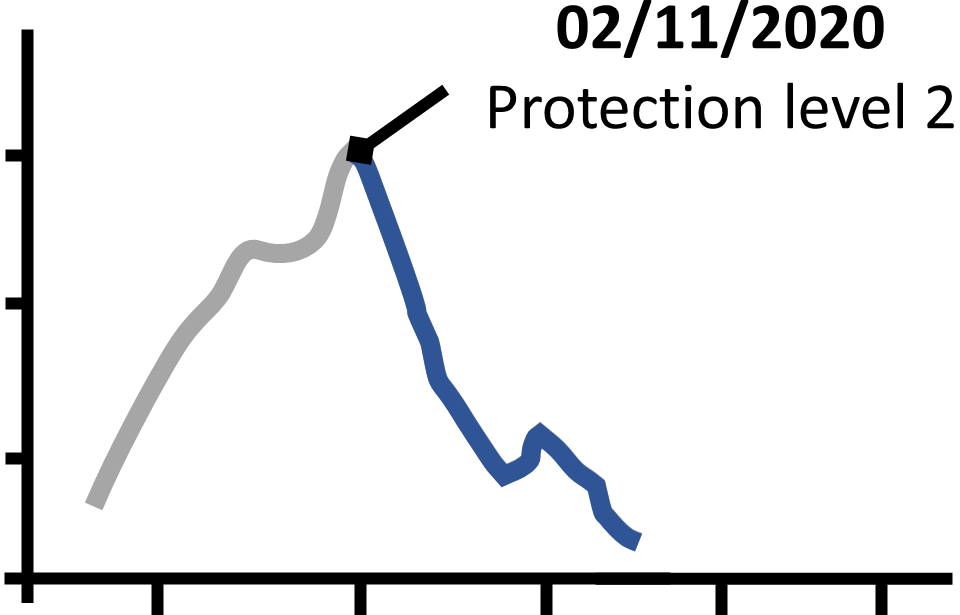}\label{SUBFIG:action3}}  \hspace{3mm}
     \subfloat[mark]{\includegraphics[width=1.5cm]{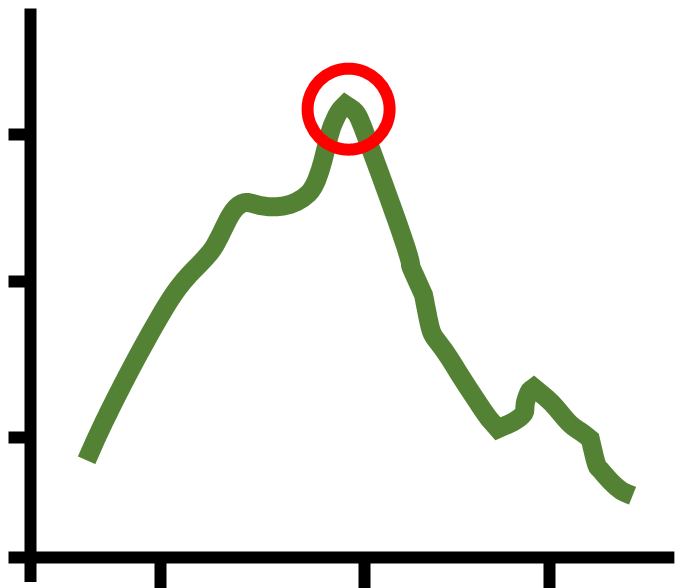}\label{SUBFIG:action2}}
     \hspace{3mm}
      \subfloat[annotate]{\includegraphics[width=2cm]{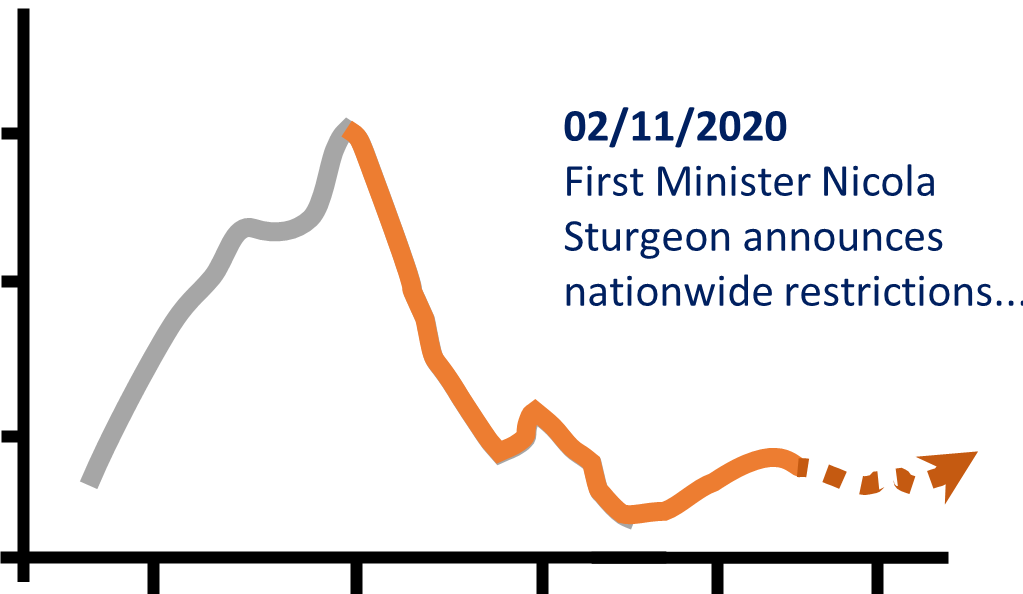}\label{SUBFIG:action4}}
    
    \caption{Actions in our implementation with their parameters: 
    (a)~change the colour of a section of the time series line for highlighting. 
    (b)~draw a circle at a datapoint. 
    (c)~annotate the graph with a line attached at a particular point. 
    (d)~\ While animating a time series segment, place a text description on the opposite half of the graph. 
    It is possible to only animate a time segment 
    or animating a segment and annotating a point.\vspace{-5mm}}
    \label{fig:Actions}
\end{figure}

Finally, the meta-author needs to consider what \textbf{actions}~\action{} occur for each encountered feature. Each action may highlight the feature visually and display a narrative (e.g., showing values and predefined text strings). A few examples of implemented actions can be seen in~\cref{fig:Actions}.
In the following sections, we outline three use cases from the RAMPVIS project, created with our meta-authoring process: (1) a story from single location with one time-series, (2) a comparison story, and (3) a scrollable time-line story.

\subsection{Single location and time Series}
\label{SUBSEC:single}

\begin{figure}[!ht]
    \centering
    \includegraphics[width=\columnwidth]{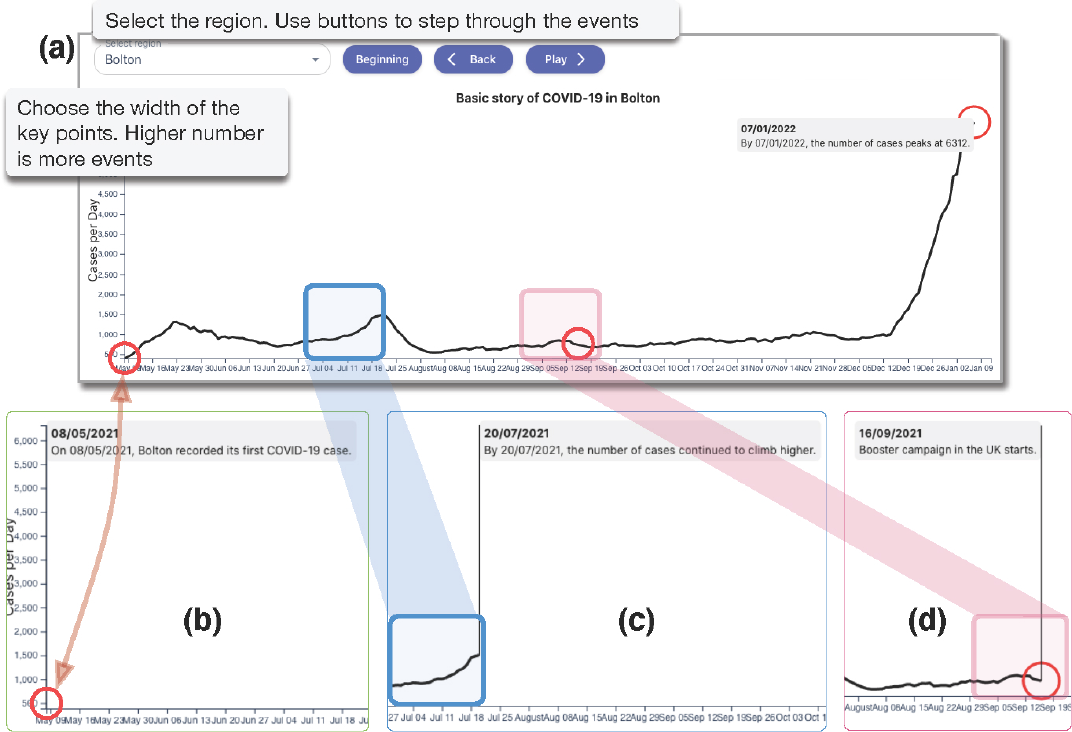}\vspace{-3mm}
    \caption{Demonstration of the single-location story. Programmed features are located. When the Play button is pressed the delivery system actions to progresses to the next feature, the appropriate text is concatenated with real data, and placed in a suitable position (if there is space it will be shown to the right of the vertical line, otherwise to the left). (a) The full interface; (b) story start, (c) highest deaths per day, (d) booster vaccination program starts. \vspace{-0.5cm}}
    \label{fig:story1}
\end{figure}

\rev{The first story (\cref{fig:story1}) focuses on one location, the quantity of positive \covid{} cases and major events such as lockdowns, vaccination program progress, etc. The story involves two time series: a)~\covid{} case data which is continuous and spread over time, and b) related events, categorised in terms of importance.}

\rev{In this case, the user can progress/rewind the story via buttons. An importance ranking value determines which events will be displayed, i.e., whether all or more significant events are displayed. The ranking of these events is decided by the author, whereas the ranking mechanism is explained in~\cref{sec:importanceRanking}.
As features are encountered and actions initiated, text descriptions are concatenated from author defined text strings, with location and data values from the data enriching the story text. This is shown in a labelled text-box linking the story text directly to the specified feature (\cref{fig:story1}(c)).}

\rev{This story's author used a spreadsheet to organise and map features and actions, which can be a suitable approach for several reasons.
Exemplar features can be added to the spreadsheet and then edited into general ones. The order of the features can be readily re-arranged, whereas different priorities that help to identify the importance and order of the story elements can also be added to the spreadsheet. 
In addition, it is simple to map features alongside actions, as different cells. Furthermore, code can be translated and aligned closely to the spreadsheet content, or even parse it.} 

\subsection{Comparative Storyboard: Two Numerical Time Series}
\label{SUBSEC:comparison}

\begin{figure}[t]
    \centering
    \includegraphics[width=\linewidth]{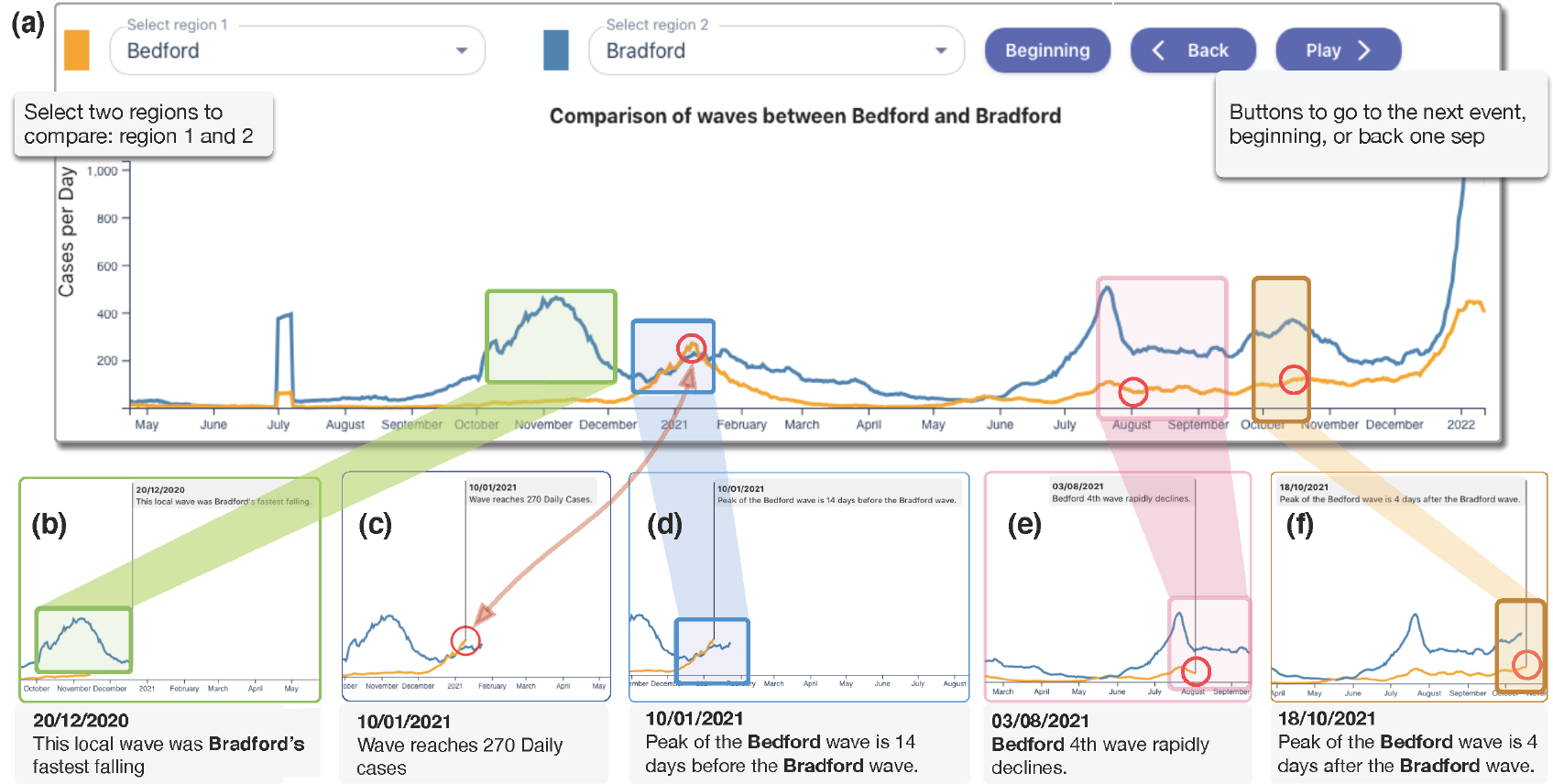}\vspace{-4mm}
    \caption{Comparison story demonstration, where (a) depicts the final frame. The story is shown in stages, moving key `features', and alternating `actions' between region 1 and 2. The insets (b-f) depict several key event features, which  are incrementally shown as the story progresses; (b) a single feature and action about Bradford (region 2); (c) story action focusing on peaks, with data specific to the local site; (d and f) comparison feature showing  differences in terms of days; (e) feature comparison based on calculated data.\vspace{-5mm}}
    \label{fig:story2}
\end{figure}

The comparison story (\cref{fig:story2}) allows viewers to compare two different locations. In our example, a user selects Bedford (100,000 inhabitants, 50 miles from London) and Bradford (350,000 inhabitants, 150 miles North of Bedford), both in England, under the same \covid{} restrictions and reasonably close; yet on different rail routes to London and distant enough to have separate ecosystems. Investigating them might indicate if the pandemic was moving towards or away from London.

\rev{Similar to the previous example, each line-graph corresponds to a location. Because there are two time series, the comparison between them is the key narrative element. The meta-author defined the same set of features to be detected for both locations. When the detected features were visualized, it appeared to alternate between features in the two line-graphs. Such features are relevant to the characteristics of \covid{} waves; e.g., cases increased in two weeks.}
\rev{The meta-author defined actions according to features. Messaging for this story has two main goals: (a) to highlight features on each time series, in a similar manner to the previous story, and (b) to highlight comparisons between the time series~\cite{GleicherETAL2011}.} 

\rev{This example explicates several key facets. First, the story starts with the region that has the first significant event; in this case, it starts with Bradford as shown in \cref{fig:story2}(b). Second the animation automatically alternates between regions and significant events.
The meta-author designed the story of each city around its major COVID-19 waves, which are identified by matching peak-related features in the data.
The meta-author instructed that events for both cities were to be grouped around the same wave and to be played in the order of waves.
With such grouping, comparisons can be made between the two cities. For example, in~\cref{fig:story2}(b) the story focuses on Bradford, and then in~\cref{fig:story2}(c) the other region (Bedford) catches up. Later, the developer implemented this instruction by processing of features in two time series concurrently. 
The meta-author also wished that the highest ranked features were shown and highlighted.
Our example depicts a few events from one region, before moving to events from the other. Regions with higher ranking, or higher data values, will play first. Third, we compare events; whether they are different, same or larger. We look at significant rises, declines and peaks; see \cref{fig:story2} (d), (e) and (f), respectively. For instance, \cref{fig:story2}(d) shows how the text explicitly compares two values (saying that Bedford was 14 days before Bradford). 
The point of interest is indicated with red circle and a vertical text label (generated from the data).} 

\subsection{Scrollable Storyboard: Two Numerical Time Series}
\label{SUBSEC:scrollable}

\begin{figure}[!ht]
    \centering
    \includegraphics[width=\columnwidth]{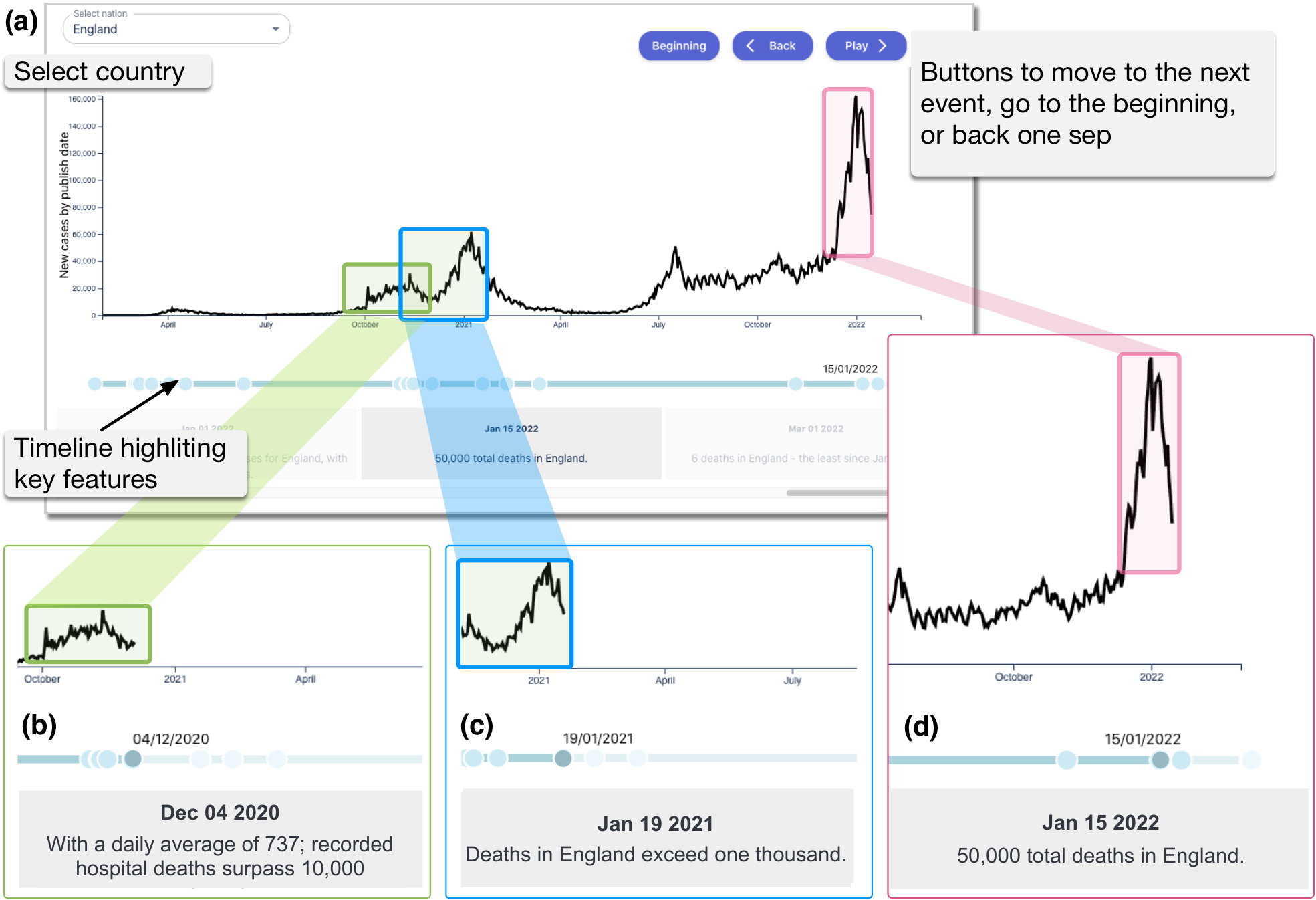}\vspace{-4mm}
    \caption{The country story (in this example focusing on England). (a) Shows is a screenshot near the end of the story. Insets (b-d) highlight three features. As the reader plays the story, so the line-graph animates to reveal the next part of the story, the text narrative animates (sliding to reveal the next text). The text narrative shows the current text bracketed by the previous and next story text.\vspace{-6mm}}
    \label{fig:story3}
\end{figure}

\rev{Our third demonstration provides a holistic scrolling story of the events in England and includes three additional facets. First we display data as a line-graph, which provides an overview of all events in the storyline. Second, we present text descriptions for the current event and elided versions for the previous and the forthcoming event. The text scrolls as the audience plays the story. In this manner, the current story text is shown in the context of what has happened, plus a hint of what is to come (\cref{fig:story3}).}

\setlength{\belowcaptionskip}{-10pt}

\rev{The scrollable story was designed with two abstractions in mind: (a) the detail of individual, and (b) the overview of all, features (cases and deaths time series). The design was captured using the FDS~\cite{Roberts_ETAL2016} method. 
Features were determined in relation to either cases, deaths, or both, and an overview is depicted in the event-line. Each highly-ranked feature is highlighted by a circle, with the current one darker. Current, previous and next event text descriptions are updated for each event.}

\rev{
Important features, such as ``first recorded case'', number of deaths or hospitalisation cases in tens of thousands, rise in cases by 20\% and so on, can be automatically detected respectively. 
By defining features with notable pre-defined values it is possible to signify interesting events from the time-series data as well. For example, people will be interested in the first case, when cases (or deaths) reach 100, 1000 and so on. 
Finally, events such as lockdown start/end dates, vaccination start dates, quotes from parliamentarians etc., can be ranked as in the previous examples.} 
\begin{figure*}[ht]
 \centering
 \vspace{5mm}
 \includegraphics[width=0.9\textwidth]{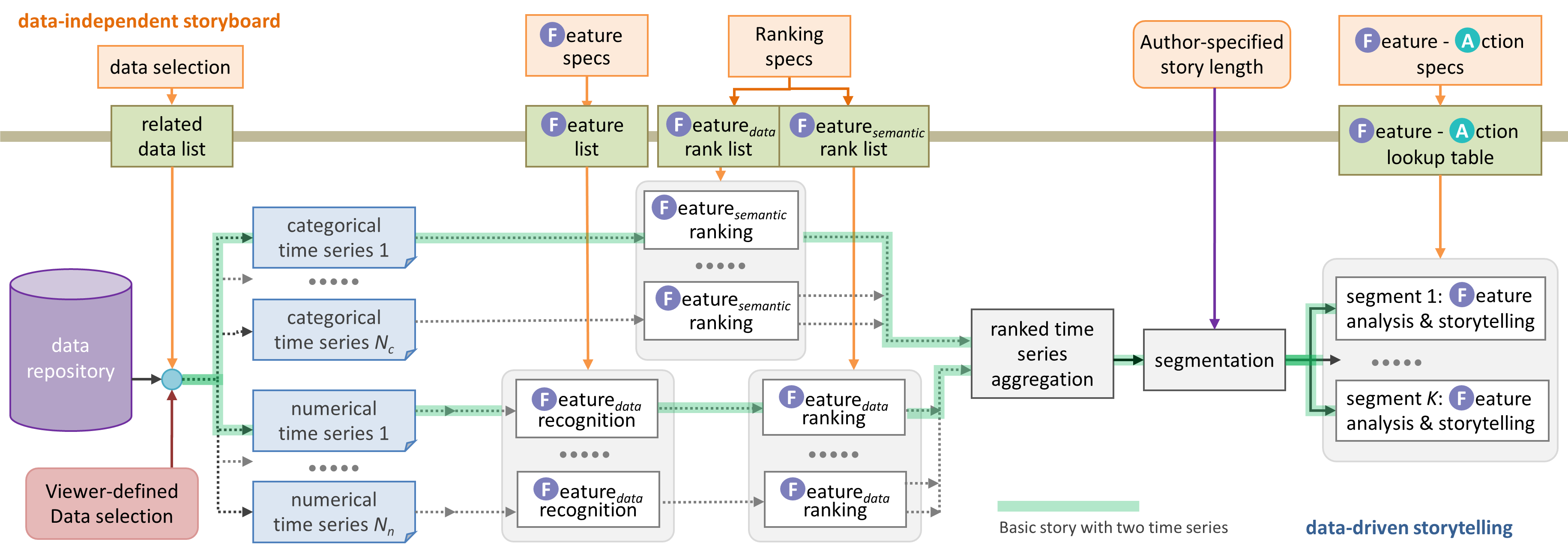}
 \caption{%
Overview of our system's workflow. The author creates a story by selecting a data which require a specific dataset from the repository. They then define features, determine the ranking (importance) of these features and map detected features \feature{} to specific actions \action{}. They also define the effective length of the story (\mySquared{workflowOrange}). These are converted by the developer in lookup tables and lists used by the algorithm (\mySquared{workflowOlive}). The lists are then used to process the different categorical and numerical time series (\mySquared{workflowBlue}), by detecting and ranking the defined features (\myRounded{workflowGray} stages). The time-series are then aggregated, and segmented based on the author-specified story length. Once the user selects the location(s) of the story~(\mySquared{workflowRed}) the story commences, and actions are invoked when any corresponding features are detected.
The path for the simple story, described in~\cref{SEC:usecases}, is highlighted in green. Further technical details can be found in Appendices A and B in the supplementary materials.
\vspace{-6mm}
}
 \label{FIG:software_arch}
\end{figure*}

\section{Algorithmic Pipeline}
\label{SEC:Algorithm}

\begin{figure*}[!ht]
    \centering
    \subfloat[A numerical time series and its detected features]{\includegraphics[width=\columnwidth]{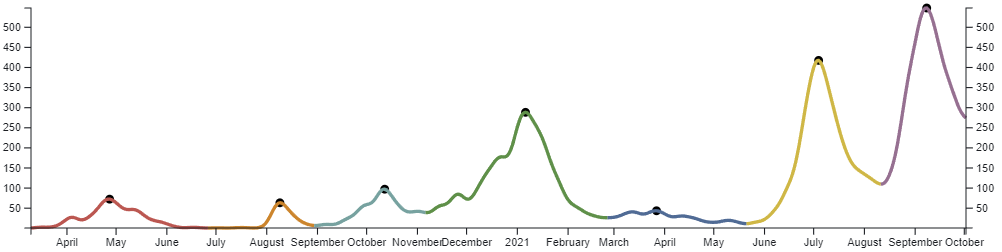}\label{SUBFIG:NTS}}\hspace{2mm}
    \subfloat[A categorical time series and its defined features]{\includegraphics[height=2cm,width=\columnwidth]{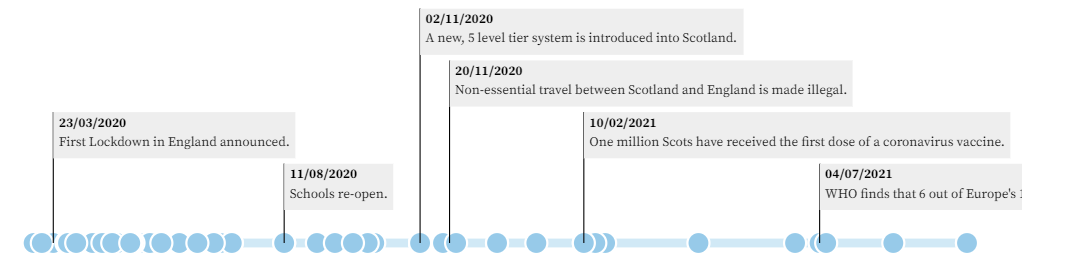}\label{SUBFIG:CTS}}
    \qquad
    \subfloat[Rankings of the features in (a) and the corresponding Gaussians]{\includegraphics[width=\columnwidth]{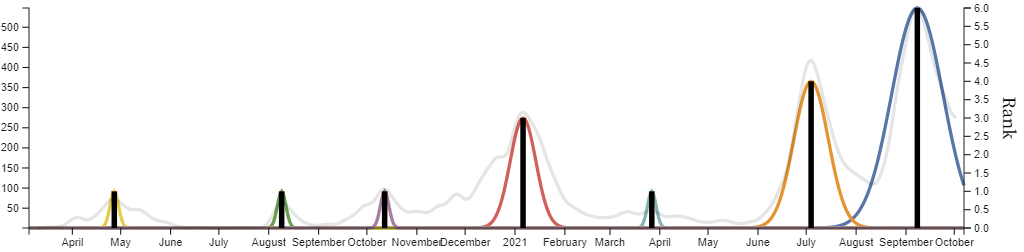}\label{SUBFIG:gaussian_ranking}}\hspace{2mm}
    \subfloat[The Gaussians of the features in (b)]{\includegraphics[width=\columnwidth]{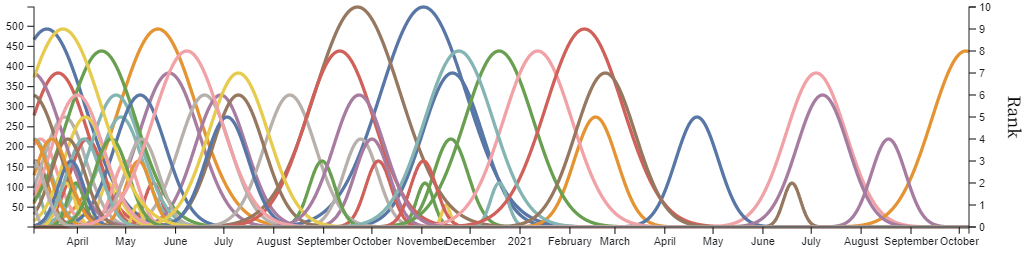}\label{SUBFIG:semantic_events}}
    \qquad
    \subfloat[Obtaining an overall importance curve using Gaussian mixture models]{\includegraphics[width=\columnwidth]{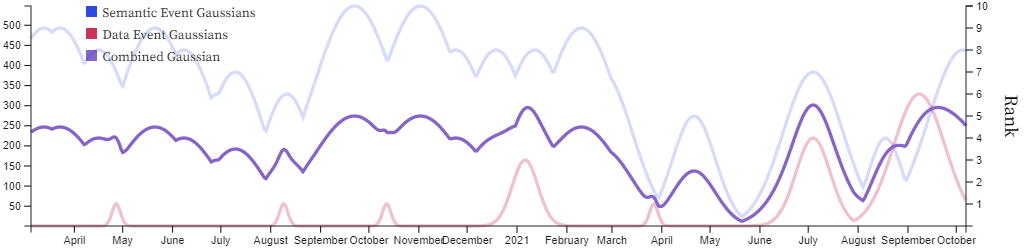}\label{SUBFIG:max_bounds}}\hspace{2mm}
    \subfloat[Segmentation based on the overall Gaussian importance curve]{\includegraphics[width=\columnwidth]{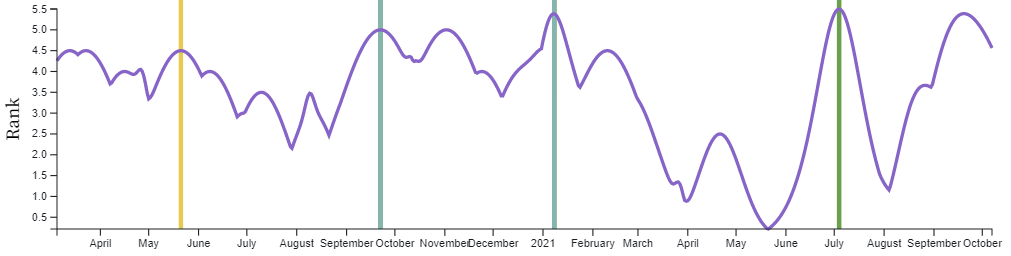}\label{SUBFIG:segmentation}}
    \qquad
    \caption{ 
    Graphical depiction of the algorithmic pipeline for automated storytelling.
    \protect\subref{SUBFIG:NTS}~Peaks, and rising and falling segments of the time series are detected. 
    \protect\subref{SUBFIG:CTS}~Event features in the categorical time series are typically predefined and ranked. 
    \protect\subref{SUBFIG:gaussian_ranking}~Each detected data feature is given a rank (illustrated as black bar). Each ranking value is converted to a Gaussian curve.  
    \protect\subref{SUBFIG:semantic_events}~Likewise, ranked semantic events are converted individually to Gaussian curves.
    \protect\subref{SUBFIG:max_bounds}~Gaussian curves are combined using Gaussian mixture models. Here a max-model is used within a time series and a mean-model is used between time series. 
    \protect\subref{SUBFIG:segmentation}~The story is divided in segments according to the combined importance curve. The number of segments is defined by a meta-author. In this example, the three green lines show the results of the segmentation algorithm, and four yellow lines indicate the segmentation results if five segments are required (three yellow lines happen to coincide with the green lines).  \vspace{-7mm}
    }
    \label{fig:Gaussian}
\end{figure*}

In this section, we describe the underlying algorithmic pipeline of our method.
Figure~\ref{FIG:software_arch} illustrates an overview of our method's workflow. The orange boxes in the top row show the human inputs, of the story viewer (data selection by virtue of story and location selection, fetched from the data repository) and the storyteller (remaining inputs). The light green boxes along the thick brown line is the data interface that translate the viewer's and storyteller's inputs to inputs algorithms can process. The remaining boxes, below the light brown line, illustrate the data flow and algorithmic process in the flow. We consider two categories of time series data:

\noindent\textbf{Numerical time series} (NTS) -- This category includes commonly-encountered time series during the \covid{} pandemic, such as daily, accumulated, normalised, and $k$-day moving average data, with semantics such as number of cases, hospitalisations, fatalities, vaccinations, etc. \cref{fig:Gaussian}(a) shows an example of a time series of the infection rate in Aberdeen City between March 2020 and October 2021. A feature detection algorithm that searches for peaks is applied to the data. Each detected peak is segmented and highlighted by a different colour and its apex indicated by a black dot.  

\noindent \textbf{Categorical time series} (CTS) -- This category extends the notion of NTS by considering each data point at time $t$ can have a categorical or nominal value. Assuming that ``null'' or ``no value'' is a valid categorical value, every data point in the time series has a value in the same way as NTS. Fig.~\ref{fig:Gaussian}(b) shows an example of a categorical time series, represented as a timeline featuring major healthcare policy changes, such as ``lockdown''. The example details events between the period of March 2020 to October 2021. Each blue dot on the timeline represents a different semantic event.

\subsection{Feature Detection}
Feature detection is a widely-used data analysis process in many applications, such as signal, speech, and image processing. In a storyboard about time series, there are some common features, such as ``peaks'', ``valleys'', ``steep climb/fall'', data milestones, smoothness of data, etc. Nevertheless, the design of a feature detection algorithm may not always be straightforward. For example, a ``peak' may be defined a local maximum point in a time series -- there are usually a lot of them. Alternatively, it may be defined a global maximum point -- there is often just one such point, or if there are a few, they are unlikely distributed sparsely in different periods. On the other hand, a story author typically requires a more complex feature definition, e.g., a local maximum that is of some distance from other peaks. Furthermore, the way a human may identify a peak is completely different to how a computer mathematically defines one. In our case, the feature detection algorithms had to go through several rounds of adjustments to match the desires of the story designers. An example of the results of our peak detection algorithm can be seen in \cref{fig:Gaussian}(a). 
The algorithm selects possible peaks from the set of all maximal points in the data. In decreasing height order, we traverse down the slopes to the left and right of each maxima until a minimum point is reached. In doing so, the bounds of the peak are found and any maxima along the way, considered part of the same peak, are removed from the selection pool. The data is segmented into different peak regions, as can be seen in \cref{fig:Gaussian}(a), with different colours representing different peaks.
In many ways, algorithms for detecting different features in an NTS assign categorical values to the data points at different time points. These categorical values collectively define a CTS.

\subsection{Importance Ranking}
\label{sec:importanceRanking}
Given a CTS with many different categorical values, a story designer will almost always wish to place different levels of emphasis on different data points according to their categorical values. For example, a ``full lockdown'' event may be considered more important than a minor policy change.
Similarly, given an NTS, following feature detection processes, the detected features can also be associated with different levels of importance.
As shown in \cref{FIG:software_arch}, we conveniently refer to the importance ranking for a CTS as \feature{}eature$_{semantic}$ ranking, and that for an NTS as \feature{}eature$_{data}$ ranking. In our system, importance ranking processes allows real-value ranking with a maximum value ($r_\text{max}>0$) that must be consistently defined within each story. For all our case studies, we defined $r_\text{max} = 10$.
The black bars in \cref{fig:Gaussian}(c) indicate the importance ranking values assigned to each peak feature detected in \cref{fig:Gaussian}(a), and ranking takes into account the height of the peak.

Because the importance ranking is based on categorical values of events and features, the ranking is somehow discrete and the interactions between different raking values (e.g., the black bars in \cref{fig:Gaussian}(c)) are difficult to compute numerically. We thus assign a \emph{Gaussian component distribution} (commonly referred to as a \emph{Gaussian} for short) to each ranking value as Gaussians indicate an interaction between two features/events and the level of interaction can be numerically computed.
In many cases, there could be many interactions, such as \cref{fig:Gaussian}(d), where each Gaussian corresponds to each categorical values (except null) in \cref{fig:Gaussian}(b).

In storytelling visualization, Gaussian mixture models were previously applied to video data  \cite{Parry2017_Snooker}.
As Gaussians are continuous curves in $[-\infty, \infty]$, we can use Gaussian mixture models to obtain a continuous importance curve by combining all Gaussians within a time series and an \emph{overall importance curve} by combining the importance curves of all NTS and CTS related to a storyboard. For example, in \cref{fig:Gaussian}(e), the light red time series is the mixed importance curve of the Gaussians in \cref{fig:Gaussian}(c), while the light blue time series is that of the Gaussians in \cref{fig:Gaussian}(d). A $\max()$ Gaussian mixture function is used in both cases. These two mixed importance curves are then mixed using a $\text{mean}()$ Gaussian mixture function, yielding the purple overall importance curve that encodes the importance features in \cref{fig:Gaussian}(a) and events in \cref{fig:Gaussian}(b).\vspace{-1mm}

\subsection{Timeline Segmentation}
Once we have obtained the overall importance curve for a storyboard, we can start the process of generating a storytelling visualization.
Similar to telling a story in speech or writing, we would like to select relatively more important features to be ``mentioned visually'' in a story according to the time or other resources available to the story.
In this work, we consider two types of resources, the \emph{number of button pressing actions} in interactive story progression, and the \emph{total animation time} in fully automated story progression.
If there were too many button pressing actions or the animation were too long, the viewers would become weary or impatient.
If there were too few interactions or the animation were too short, the viewers may find the information provided inadequate.
Technically, we would like to divide the timeline of a storyboard into $k$ sections. In interactive story progression, $k$ sections require $k$ button pressing actions (including the start). In automated story progression, $k$ determines the total animation time as $k \times$``unit section time''.
\cref{fig:Gaussian}(f) shows two possible segmentation results with $k=3$ and $k=5$ respectively. 

We developed an algorithm to segment the timeline of a storyboard by considering the peaks in the overall importance curve of the storyboard. The algorithm selects the top $k-1$ peaks with a maximal gap $\Delta_\text{max}$ between neighbouring peaks.
In this way, we ensure that the most important temporal points are selected and located at the section boundaries, where we can insert more visualization actions. 
With interactive story progression, users can spend more time to view the current visualization before pressing the button for progression. In automated story progression, we can slow the animation at each boundary between the two sections.   

\subsection{Feature-Action Profiling}

As aforementioned, the output of the meta-authoring process is a feature-action specifications expressed by the story author in different ways (e.g., textual descriptions, pseudocode and sketches). A developer translates the specifications to feature-action data patterns in a lookup table, where \emph{features} are categorical labels of all features that might be detected in the NTS and CTS select by story viewers, and \emph{actions} are calls to appropriate functions of the software components, for displaying various visual artefacts (\cref{fig:Actions}) as well as exhibit the desired software behaviours in terms of interaction, graphics, and animation.

The story's length is controlled by how many segments the Gaussian curve is split into and how many features there are in the lookup table. A story designer can extend the time a story takes to complete, by increasing the number of features in the table to pick up on. This timing can also be controlled using the feature rankings. For instance, to decrease the time to complete a story one could restrict the algorithm to showing only the most important events, say the top three events in a segment or maybe just those with a ranking of 8 or greater.

The algorithm for feature-action profiling visits each segment one-by-one. Within segments each feature is visited in chronological order.
Each time, it refers to the feature-action lookup table to check if the current feature matches any entry. If so the visual action is initiated.\vspace{-1mm} 

\subsection{Generation of Storytelling Visualization}
After all sections have been profiled, the ``play'' button becomes available to users. In the mode of interactive story progression, a viewer presses the button to activate the storytelling visualization of a section.
In response, the underlying algorithm visits the NTS and CTS of user-selected region in the order of time. For any feature (or event) that is marked as being included in the story, the algorithm invokes the corresponding action. When the end of the section is encountered, the algorithm waits until the viewer presses the ``play'' button, to proceed to the next section.
In the mode of automated story progression, instead of waiting for a viewer's interaction, the algorithm waits for a short time span, and continues to the succeeding section automatically. To make all stories have a consistent look and feel, this mode is not currently deployed.\vspace{-1mm}

\section{Evaluation and Reflection}
\label{SEC:evaluation}

To evaluate our approach, we grouped stakeholders of this work according three perspectives: (i) story authors, (ii) software developers, and (iii) public members and one public engagement expert. The first two roles are the intended users of our approach, where as the latter were chosen to provide insight on the viewer-centric nature of our method's outcomes.
\rev{We requested their written feedback, using a shared online document and collected their comments as evaluation and reflection, similar to a retrospective verbal protocol analysis.}  While we provided five questions for each group as prompts, we gave explicitly the direction ``please feel free to add any comment.'' The full text of the prompts and feedback is given in Appendix C in the supplementary materials.
The main feedback points are summarised below, with direct extractions from stakeholders' input in \emph{italics}.

\subsection{From the Perspective of Story Authors}
 
Three story authors (who created the stories in \cref{SEC:usecases}) provided feedback of more than 1400 words. The main points are summarised below:

\begin{itemize}[leftmargin=10pt,noitemsep,topsep=0pt]
    \item Meta-authoring can be more challenging than telling a story about a single dataset. Authors \emph{need to think about the story holistically}, and how different data, say from different regions, may \emph{support key messages}. \emph{It forces authors to consider the underlying data in detail,} which is \emph{a good practice in storytelling}.
   
    \emph{Generalizing the story is definitely a gain from using this approach.} \emph{Creating a draft storyline helped think about how to generalise and abstract the information into the meta-instructions}.
    \item Creating feature-action design patterns can be \emph{somewhat challenging}.
    \emph{Certainly some of the key features were obvious, such as a peak, rise and fall. But not all were easy} to define.
    It seems to make an author consider \emph{the progression of the story in a more ``mathematical'' way}. \emph{Information needs to be abstracted or generalised into instructions} for software developers.
    
    \item \emph{The workflow relies on the software engineer to both interpret the design and create code}. It is appropriate at the moment with the state of the art technology for storytelling visualization, but \emph{there are opportunities for further automation}. The approach of having \emph{software engineer in-the-loop} may work for \emph{large organisations}, but independent meta-authors or those in \emph{small organisations} can benefit from meta-authoring tools.  
    \item In the future, \emph{an advanced interface could be developed} for meta-authors \emph{to create these feature-action} design patterns. \emph{One could envision an interface where storytellers can pick and match story segments and combine them} sequentially or hierarchically.
    Feature-action \emph{abstraction can be exploited to create a more modular approach for the end users} (i.e.,  meta-authors).
\end{itemize}

\subsection{From the Perspective of the Developer}

Two research software developers were directly involved in developing the pipeline in \cref{FIG:software_arch}, various feature detection algorithms, and visualization components. They, together with a third stakeholder who helped design the algorithmic pipeline, provided feedback totalling more than 1200 words. The main points include:

\begin{itemize}[leftmargin=10pt,noitemsep,topsep=0pt]
    \item It is not difficult for a technical developer to support the meta-authors in general. Given a storyboard and visualization guidance from the meta-authors, \emph{a programmer can create the storytelling visualization easily}. \emph{The notion of feature-action design patterns was accepted by both meta-authors and technical developers.} The technical developers were able to \emph{translate} meta-authors' \emph{qualitative specification of design patterns} to the actual implementation in a \emph{look-up table relatively easily.} 
    \item \emph{An implication of creating stories for the public is that features must understandable and simple.} \emph{Developing such feature detection algorithms varied from taking a few hours to a few days.} 
    \item Regarding software components that implement actions, while it is not hard to implement them individually, \emph{it can take up to a week to} put them together for a story. Among our stories, \emph{the longest development, for an entire story took roughly four days.}
    \item \emph{Observable is a very good ``drafting'' software. It would take less than a day to implement a story in Observable if the story requires no new feature or action.} However, it was not feasible for us to place many dynamic data streams on Observable. We had to develop and deploy storytelling visualization on an infrastructure that keeps the dynamic data streams.
    \item \emph{the visualization software components (e.g., multi-line time series plots, textual annotations, timelines, animation, etc.) are implemented using D3.js}~\cite{d3_2011}. With the existing software code and programming experience, \emph{developing new software components using D3.js had been quick and easy.}
    These software components \emph{were implemented as reusable classes and functions.} Gradually, the time for implementing new software components for a new story  \emph{can be reduced} significantly. 
    \item In the future, it will be desirable and feasible to develop software systems to support the workflow from meta-authoring to story deployment. Developers and meta-authors have suggested a number of options, including
    (i) \emph{visual programming environment},
    (ii) \emph{object-oriented/template-based programming environment},
    (iv) \emph{scripting language},
    (v) \emph{markup language (e.g., XML)},
    (vi) \emph{declarative data interchange format (e.g., JSON, YAML)}, and
    (vii) \emph{form-like webpage}.
    \emph{the software components developed in this project can provide a good basis for future developments.} 
\end{itemize}

\subsection{From the Perspective of Public Engagement}
We were interested in hearing from those stakeholders who had expertise in public engagement as well as who are the potential viewers of storytelling visualization. 
One expert of public engagement provided feedback directly, and two members of the public provided verbal feedback, which were transcribed by a co-author.
Together the three stakeholders provided feedback totalling more than 900 words. The main points in the feedback include:

\begin{itemize}[leftmargin=10pt,noitemsep,topsep=0pt]
    \item \emph{There has been a surfeit of data relating to COVID-19 in the public domain, which may well seem like information overload to members of the public. Without context, a time series may not convey much meaning. The story telling both provides context and divides the data into digestible segments, allowing the viewer to look at the situation leading up to an event and following after it, and to understand/reflect on each segment before moving on to the next. ... Because the stories are told linearly and from a local area perspective, the user can situate themselves both in place and time in relation to the events, which adds to the human interest and relatability of the data.}
    \item \emph{Because it takes only one selection button for a region and then the play button to start, it is actually slightly easier than many search interfaces.} \emph{If it were available some time in 2020 (and perhaps the early part of 2021), it would get used a lot.}
    \item In the stories, \emph{the event descriptions are simple and effective, giving clear information about key landmarks. The speed of the animation provides momentum, and the event pauses give the user control over how fast to move on with the story.}
    \item \emph{The stories are likely to appeal to members of the public who are regular news viewers, who listen to scientific podcasts, and who take an interest in the local or regional context of the pandemic and how policies or actions affect its progress.}
    
    \item \emph{This form of location-dependent, data-driven storytelling would be useful and of interest to members of the public}, and could be used to visualise \emph{housing, energy and fuel prices}, \emph{school performance}, \emph{crime rate}, and \emph{transport statistics}. 
    \emph{Comparison of different cities, areas, or regions would be useful}.%
    \item \emph{Would it be possible for an individual to create such a story to share with friends? Some people are quite keen to share data, such as their daily walking step counts.}  
\end{itemize}

\section{Further Application and Evaluation}
\label{sec:ML}
While the aforementioned meta-authoring method for creating storytelling visualization was developed during the \covid~pandemic, we have always been keen to apply the method to other domains, and demonstrate its generalisability. An opportunity arose when one of our visualization (VIS hereafter) researcher joined a ML team in the Science and Technology Facilities Council (STFC) in the UK. The ML team has been providing physics, nuclear, space, and astronomy scientists with ML models for processing a variety of data captured using different sensory modalities (e.g., \cite{Hey:2020:PTRSA}). One R\&D strand of the team has been to develop and analyse ML benchmarks as the basis for providing guidelines and best practices in AI for Science \cite{Henghes:2021:MNRAS,Thiyagalingam:2022:NRP}.

\begin{figure*}[ht]
    \centering
    \begin{minipage}{.4\textwidth}
    \subfloat[provenance story]{\includegraphics[width=\columnwidth]{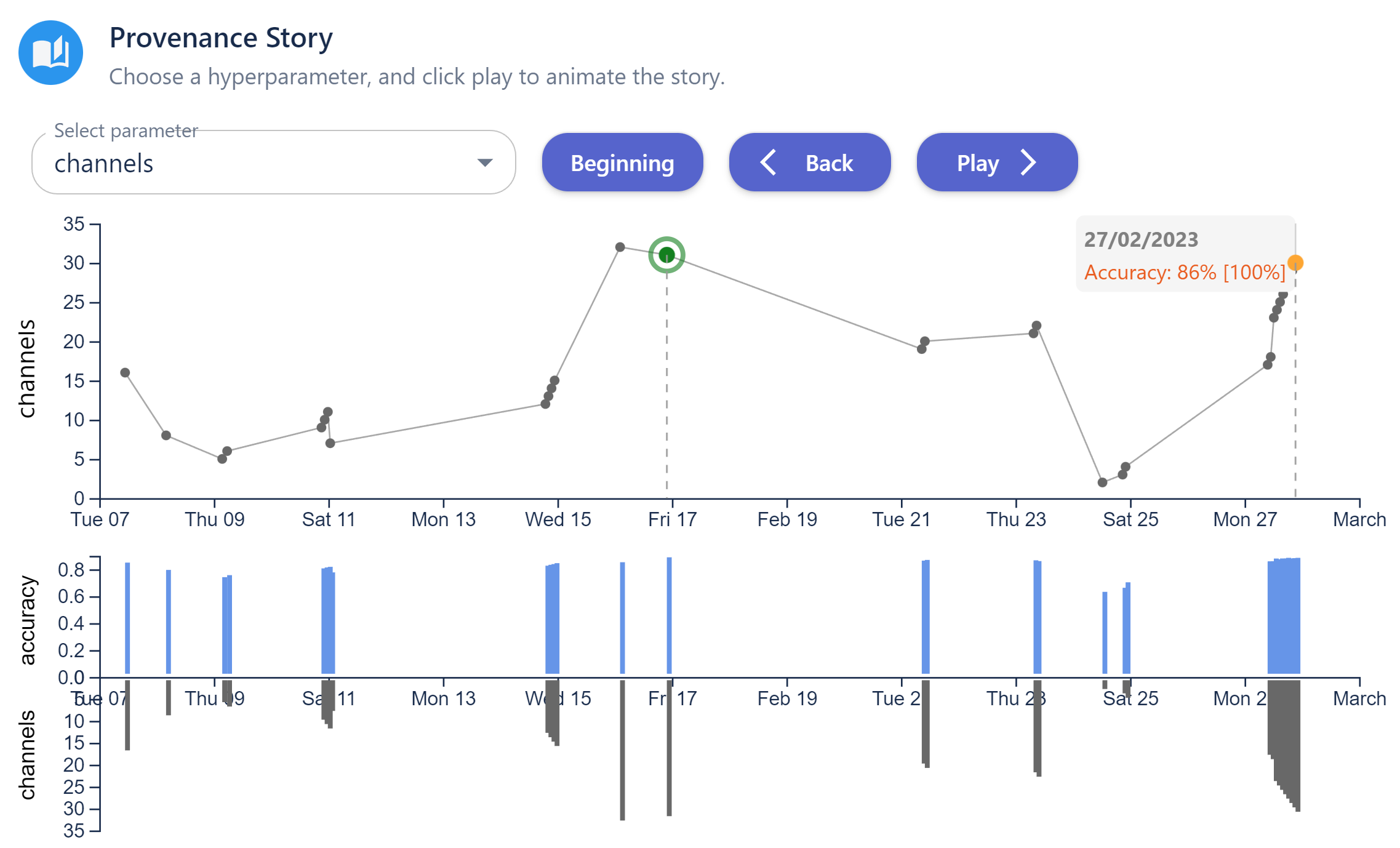}}\label{SUBFIG:ml1}\\
    \subfloat[analytical dashboard \& story]{\includegraphics[width=\columnwidth]{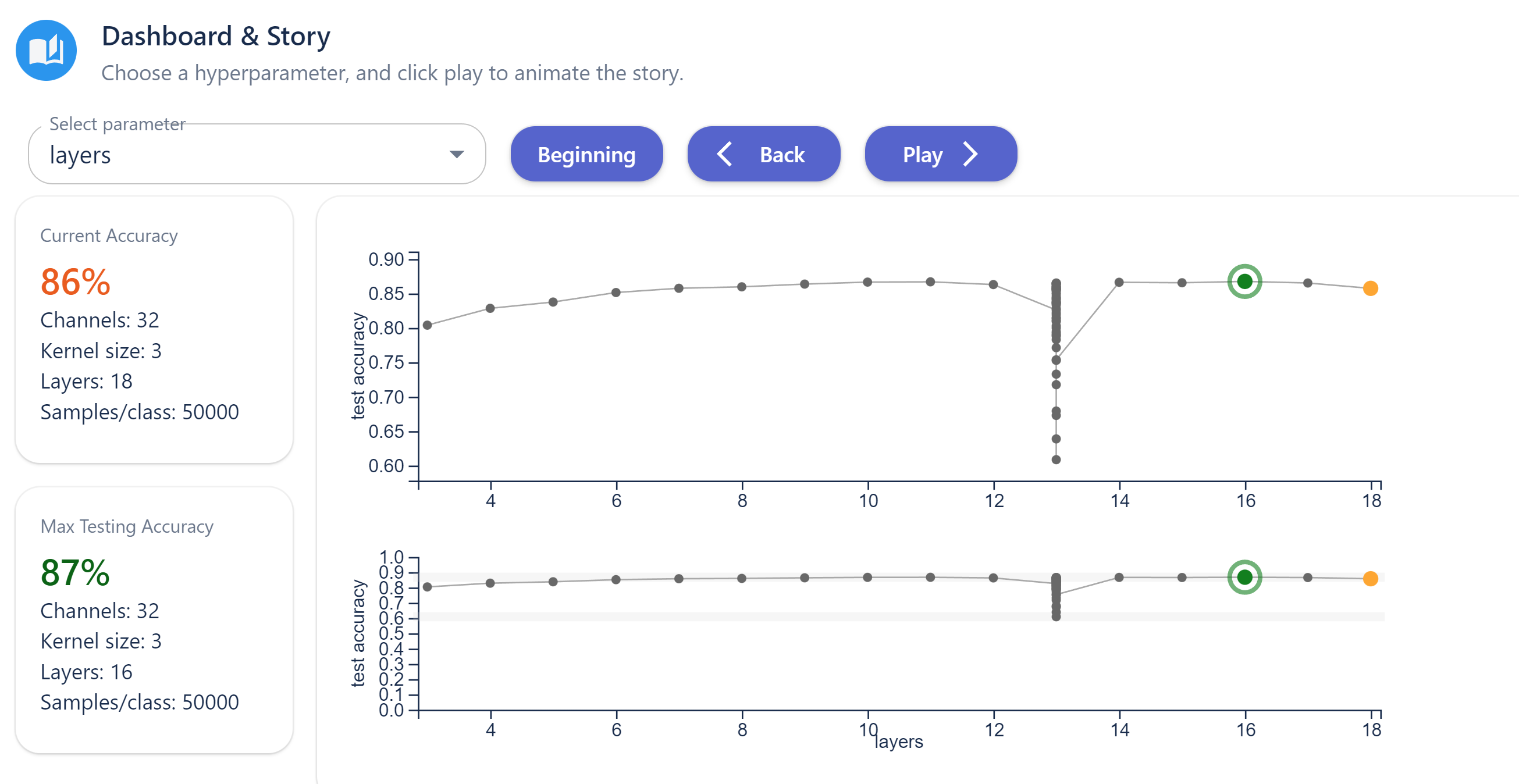}}\label{SUBFIG:ml2}
    \end{minipage}
    \begin{minipage}{.55\textwidth}
        \subfloat[multivariate story]{\includegraphics[width=\columnwidth]{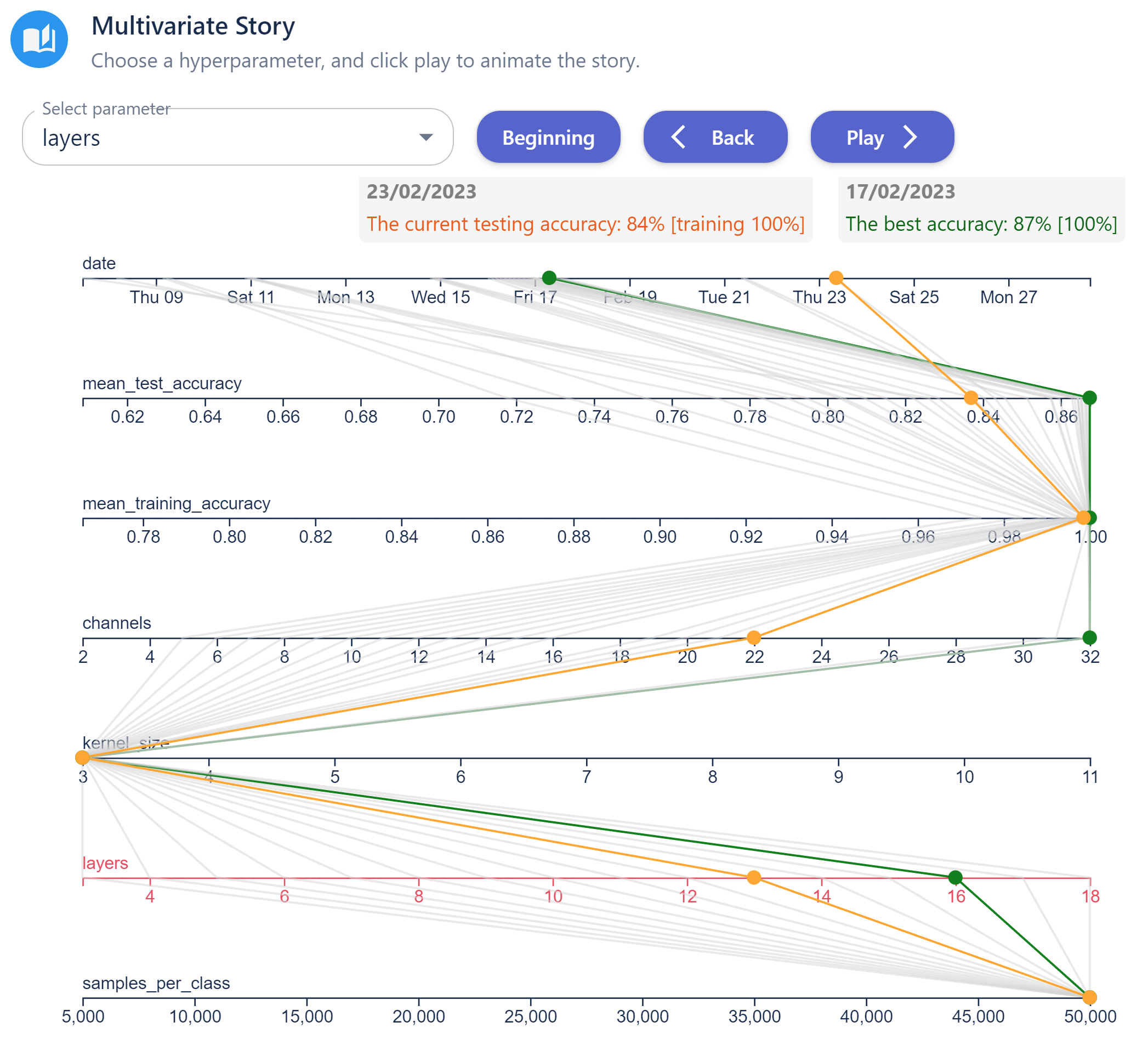}}\label{SUBFIG:ml3}
    \end{minipage}
    \caption{Screenshots of three stories implemented for supporting ML workflows developed using our approach. }\vspace{-5mm}
    \label{fig:MLworkflows}
\end{figure*}

\noindent \textbf{Requirements Analysis.} Two VIS researchers met the ML team leader and identified general requirements for using automated visualization tools in ML workflows, as well as several specific requirements for using storytelling visualization. As the team always has a few ongoing ML workflows concurrently and each may last up to 18 months, they need to observe their progress and provenance frequently, preferably though dynamically-updated stories. Following this meeting, one VIS researcher collaborated closely with an ML researcher working on ML benchmarks.
The VIS researchers designed and implemented two storyboards initially as shown in \cref{fig:MLworkflows}(a,c). During one evaluation meeting, we identified the need for an analytical dashboard with storytelling on demand. This became a new storyboard, shown in \cref{fig:MLworkflows}(b).

\noindent \textbf{Meta-Storyboards.} The STFC ML team has an established process for storing log data for ML training and testing. When a model is tested, each logged event consists of data such as date, time, hyperparameters, and accuracy measures. To re-purpose the meta-authoring software developed in the context of \covid, we focused on log data featuring time series. 

A \textbf{Provenance Story} (\cref{fig:MLworkflows}(a)) was designed around the provenance and progress of an ML workflow. A viewer can select a story to be told from the perspective of a specific hyperparameter. The animated visualization shows the logged chronological events focuses on those related to the selected hyperparameter. Interesting data features include, for instance, when the accuracy has a significant improvement or reaches a peak point. Visual actions (e.g., text message, colour highlighting) are triggered by such features.

An \textbf{Analytical Dashboard \& Story} (\cref{fig:MLworkflows}(b)) enables individual ML researchers to visualise logged events in both static and storytelling manner. Whenever a model-testing process is completed, the ML researcher concerned needs to observe the new logged event. To prevent the storytelling mechanism from slowing down the routine VIS tasks, the storyboard makes static visualization as default, with a visual design that focuses on the ordered values of a selected hyperparameter rather than event date and time. Drawing our experience of working on \covid~data \cite{Khan:2022:TVCG,Bach:2023:TVCG,Khan:2022:TSC}, we introduced elements of dashboard design to improve the efficiency of the routine VIS tasks. The storytelling mechanism can be activated when the ML researcher wishes to be reminded of the provenience, or other team members wish to be told stories about individual hyperparameters.    

Finally, a \textbf{Multivariate Story} (\cref{fig:MLworkflows}(c)) tells stories about multiple hyperparameters through a parallel coordinates plot (PCP). Viewers can select a specific hyperparameter to inform the importance ranking of features (see Section \ref{sec:importanceRanking}). The animated visualization shows the logged events chronologically, but unlike \cref{fig:MLworkflows}(a), viewers can see all hyperparameters in the same visualization. As the visualization is more complex, the feature-action patterns involve more frequent pause-and-continue actions.  

\setlength{\belowcaptionskip}{-1cm}

\noindent \textbf{Evaluation and Improvement.} In the STFC ML team, the VIS researcher was in contact with ML colleagues regularly, discussing requirements and design options while conducting frequent bite-sized evaluations. In addition, there were two formal evaluation meetings. In the first meeting, ML researchers were shown a first-draft working version of the Provenance and Analytics Dashboard stories (\cref{fig:MLworkflows}(a,c)), which prompted them to make a number of revision suggestions. We also noticed that different team members suggested a revision on the requirements of the Provenance Story. Moreover, the team wanted to observe the progress and provenance of all ML workflows in a single storytelling visualization. Finally, individual ML developers wanted to observe new results quickly whenever a trained model has been tested and observe the provenance data on demand. We decided to create a separate storyboard for these, i.e., the Analytical Dashboard \& Story (\cref{fig:MLworkflows}(b)).

Although that meeting was the first time when the ML researchers worked with a PCP, without much explanation by VIS researchers, they found the Multivariate Story particularly useful for observing the combined impact caused by different hyperparameters. This was somewhat unexpected, since many visualization students and users often encounter difficulties in learning to work with PCPs. We hypothesised that because storytelling shows data-events line-by-line, this may make the concept of PCPs more intuitive. Animation might also reduce the confusion caused by cluttering, while enabling viewers to pay attention to the interesting data-events without much visual searching effort.
In the second evaluation meeting, we focused on the Analytical Dashboard \& Story (\cref{fig:MLworkflows}(b)). The ML experts were enthusiastic about the dashboard and discussed the ideal data to be shown in the information boxes, and the interesting features in the plot, so that the storytelling mechanism highlights them automatically.\vspace{-0.8em}

\section{Conclusion}
\label{SEC:conclusion}
In this work, we presented a new method for creating storytelling visualizations, based on a workflow for meta-authoring with feature-action design patterns.
We have demonstrated the feasibility of this new method through three storyboards developed in the \covid{} visualization context, followed by three storyboards describing ML workflows. 
Our work addresses the challenge of creating generic storyboards that can be applied to different yet similar, often partially unknown data streams and can respond to different (data) features automatically using different (visualization) actions.
As summarised in \cref{SEC:evaluation} and \cref{sec:ML}, the method has been welcomed by different stakeholders, who have already looked ahead to the potential applications in the future as well as more advanced technologies that can provide system-level support to meta-authoring workflows.
Building on this work, we hope to continue improve techniques for storytelling visualization.\vspace{-0.8em}

\section*{Acknowledgments}
The authors would like to express their gratitude to the SCRC management team for its leadership, all SCRC members (including VIS volunteers) for their selfless efforts, and the UK Science and Technology Facilities Council (STFC) for providing and managing the RAMPVIS hardware platform.
In particular, we thank all other VIS volunteers for their contributions to various RAMPVIS activities since May 2020. This work was supported in part by the UKRI/EPSRC grants EP/V054236/1 and EP/X029557/1.

\balance

\bibliographystyle{eg-alpha-doi}  
\bibliography{main}        
            
\newpage

\appendix

\noindent\Large\textbf{\textcolor{gray}{Supplementary Materials}}

\noindent 
\Large\textbf{Feature-Action Design Patterns
for Storytelling\vspace{1mm}\\
Visualizations with Time Series Data}
\normalsize

\noindent
S. Khan, S. Jones, B. Bach, J. Cha, M. Chen, J. Meikle,\\
J. C. Roberts, J. Thiyagalingam, J. Wood, P. D. Ritsos


\vspace{5mm}
\noindent\textbf{Introduction to Appendices}

These set of appendices provide further details about the technical aspects of the method that is described in Section \ref{SEC:Algorithm} and illustrated in Figure \ref{FIG:software_arch}, including:
\begin{itemize}
    \item Appendix \ref{App:Feature-Action}: The general structure of a feature-action table illustrated in the top-right of Figure \ref{FIG:software_arch}.
    \item Appendix \ref{App:Meta-Storyboard}: The general structure of a meta-storyboard program for the algorithmic workflow illustrated in the lower-half of Figure \ref{FIG:software_arch}.
\end{itemize}

\section{\textbf{Feature Action Tables}}
\label{App:Feature-Action}
As shown in Figure \ref{FIG:metaprocess} (the 2nd box) and Figure \ref{FIG:software_arch} (top-right), a meta-author formulates the specification of a set of feature-action patterns, and then discusses the specification with a developer, who defines a feature-action table that can be read by a program. Normally, the program encodes a meta-storyboard and is used to create storytelling visualization.

The meta-author's specification can be documented in many forms, such as word files, spreadsheets, hand-written and hand-drawn notes, and so on, while the \textbf{feature-action table}, defined by the developer, must be in a format that the meta-storyboard program can read. We currently use CSV files for feature-action tables.
Each feature-action table has the following columns:
\begin{enumerate}
    \item \textbf{TimeSeriesId}: \texttt{<id>} of a time series in the storyboard.
    \item \textbf{Feature}: \texttt{<name>} of a feature extraction function in the Meta-Storyboard Features API (MSB Features).
    \item \textbf{FeatureParams}: a list of parameters for the feature concerned in the form of \texttt{<parameter name>:<parameter value>, <parameter name>:<parameter value>, ...}.
    \item \textbf{Rank}: a rank value $\in [1, 10]$ assigned to the action for the detected feature. 
    \item \textbf{Action}: \texttt{<name>} of an action function in the Meta-Storyboard Actions API (MSB Actions).
    \item \textbf{ActionParams}: a list of parameters for the action concerned in the form of \texttt{<parameter name>:<parameter value>, <parameter name>:<parameter value>, ...}.
    \item \textbf{Text}: a text section in the form of \texttt{<text>} defined as string template. Note that the list of parameters for the action is in the column \textbf{ActionParams}.
    \item \textbf{Comments}: optional comments by the meta-author or the developer. These are not designed to be understood by computer programs.  
\end{enumerate}

A meta-authoring storyboard may need to deal with multiple time series, e.g., a categorical time series for dates and events, two numerical time series for the number of cases in two different regions, and the ``difference time series'' derived from the numerical time series. Features can be defined for each time series independently. The current version of the software assumes that complex features for characterizing interactions among two or more time series are detected in two ways: (a) through a derived time series resulting from a subroutine that processes two or more time series, such as a difference function, (b) through the Gaussian mixture process following the feature extraction step.

\subsection{A.1 Meta-Storyboard Features API}
\label{sec:msb-features-api}

Conceptually, features are defined as functions in the MSB Features API, and the function names and their parameter names are used as predefined constants in feature-action tables. Note that developers may include undefined feature names or parameter names in a feature-action table during their communication with meta-authors with the knowledge that they will add such features and/or parameters into the MSB Features API. Meanwhile, all our storyboard programs are able to ignore undefined features and parameters in a way similar to web browsers ignoring unknown HTML tags.

In order to implement and manage similar feature functions in an object-oriented manner, we group similar feature functions using a high-level feature construct. Table~\ref{table:feature-name} shows a list of high-level feature constructs. Each construct is defined in conjunction with a list of parameters, such as (\texttt{GTE:1000, LTE:2000}) for a value range [1000, 2000]. This allows high-level constructs to be decomposed into low-level constructs. For example, the PEAK feature has low-level constructs for detecting, e.g., the $k$-th peak or a peak with additional slope criterion. For each construct, a pre-defined function, \texttt{detectFeatures()}, analyzes the given time series according to the feature parameters in the feature-action table. 

\begin{table}[htb]
    \centering
    \caption{The main features implemented in the MSB Features API for specifying time series features.}
    \label{table:feature-name}
    \begin{tabular}{@{}l@{\hspace{2mm}}l@{}}
        \hline
        \textbf{Feature Name} & \textbf{Description} \\
        \hline
        \multicolumn{2}{c}{\textbf{\textit{numerical features}}} \\
        FIRST & encountering the first data point \\
        CURRENT & setting action location at the current data point \\
        SEARCH & setting a search range (default: to the end) \\
        LAST & encountering the last data point \\
        MIN & encountering the MIN value \\
        MAX & encountering the MAX value \\
        VALUE & encountering a segment with specified value range \\
        STDEV & encountering a segment with specified STDEV range \\
        PEAK & encountering a peak (not necessarily MAX) \\
        VALLEY & encountering a valley (not necessarily MIN) \\
        RISE & encountering a rising segment \\
        FALL & encountering a falling segment \\
        SLOPE & encountering a segment with specified slope range \\
        \hline
        \multicolumn{2}{c}{\textbf{\textit{categorical features}}} \\
        EVENT & encountering an event with a specific event label \\
        \hline
    \end{tabular}
\end{table}

\subsection{A.2 Meta-Storyboard Actions API}

Similar to features, actions are conceptually defined as functions in the MSB Actions API, and the function names and their parameter names are used as predefined constants in feature-action tables. Similarly, all our storyboard programs are able to ignore undefined actions and parameters.
Similar action functions are also grouped together using a high-level feature construct. Table~\ref{table:actions-name} shows a list of high-level action constructs. Action parameters facilitate the decomposition of a high-level construct into low-level constructs.

When a feature is detected, the corresponding action is not activated immediately in order for all feature-action patterns to be further processed (e.g., Gaussian mixture, segmentation, and rank-based event selection). Hence when a feature is detected, a function \texttt{registerActions()} registers an action at the data point where the feature is encountered.   

\begin{table}[htb]
    \centering
    \caption{The main actions implemented in the MSB Actions API for specifying visualization actions.}
    \label{table:actions-name}
    \begin{tabular}{@{}l@{\hspace{2mm}}l@{}}
        \hline
        \textbf{Action Name} & \textbf{Description} \\
        \hline
        DRAW\_DATA & display a set of data points \\
        DRAW\_AXIS & display an axis or all axes \\
        TEXT\_BOX & display a text string in a predefined message box \\
        TEXT\_POS & draw or remove a text string on the plotting canvas \\
        LINE & draw or remove a highlighting line \\
        CIRCLE & draw or remove a highlighting circle \\
        RECTANGLE & draw or remove a highlighting rectangle \\
        ARROW & draw or remove a highlighting arrow \\
        NTS & highlight or dehighlight a numerical TS segment \\
        CTS & highlight or dehighlight a categorical TS segment \\
        NODE & highlight or dehighlight a graph node \\
        CONNECTOR & highlight or dehighlight a graph edge (connector) \\
        AXIS & highlight, or dehighlight a section of an axis \\
        PAUSE & pause the animation for a specific amount of time \\
        \hline
    \end{tabular}
\end{table}


\subsection{A.3 Example Feature-Action Tables}

We present two example feature-action tables, Tables~\ref{table:ex-feature-action} and ~\ref{table:ML-feature-action}. They are for illustration and they are simpler than the meta-storyboards that were developed in our case studies. 

\begin{table*}[t]
    \centering
    \caption{An example of a feature action table used to create a single-region meta-storyboard with COVID-19 data. The visual results are similar to the screens shown in Figures~\ref{fig:story1} and \ref{fig:story3}. Empty cells are interpreted as null values.}
    \label{table:ex-feature-action}
    \begin{tabular}{|>{\tiny}m{0.4cm}
                    |>{\tiny}m{1cm} 
                    |>{\tiny}m{1.8cm} 
                    |>{\tiny}m{0.4cm}
                    |>{\tiny}m{1.4cm}
                    |>{\tiny}m{3.1cm}
                    |>{\tiny}m{4.5cm}
                    |>{\tiny}m{1.6cm}|}
        \hline
        \textbf{ID} &
        \textbf{Feature} & 
        \textbf{Feature Parameters} & 
        \textbf{Rank} & 
        \textbf{Action} & 
        \textbf{Action Parameters} & 
        \textbf{Text} &
        \textbf{Comment}
        \\ \hline
        TS1 & FIRST &      & 10 & DRAW\_AXIS & & &
        \\ \hline
        TS1 & VALUE & GT:0 & 5 & DRAW\_DATA & & & $>0$
        \\ \hline
        TS1 & CURRENT &    & 5 & TEXT\_BOX & BOX:1 & \{REGION\} recorded its first COVID-19 case. &
        \\ \hline
        TS1 & SEARCH & UPTO:28 & 10 & DRAW\_DATA & & & up to 28 days
        \\ \hline
        TS1 & RISE & SLOPE\_GTE:15 & 5 & TEXT\_BOX & BOX:1 & The number of cases grew. & \\ \hline
        TS1 & SEARCH & UPTO:28 & 10 & DRAW\_DATA & & & up to 28 days
        \\ \hline
        TS1 & SLOPE & GTE:10 & 6 & TEXT\_BOX &  & By \{DATE\}, the number of cases continued to climb higher. Let us all make a great effort to help bring the number down. Be safe and support the NHS. & case A
        \\ \hline
        TS1 & SLOPE & LTE:-10 & 6 & TEXT\_BOX &  & By \{DATE\}, the number of cases dropped noticeably. Excellent effort. Be safe and support the NHS. & case B
        \\ \hline
        TS1 & SLOPE & GT:-10, LT:10 & 3 & TEXT\_BOX &  & By \{DATE\}, the number of cases remained low. We should continue to be vigilant. & case C
        \\ \hline
        TS1 & PEAK &   & 10 & DRAW\_DATA & & & default parameters
        \\ \hline
        TS1 & CURRENT & & 10 & TEXT\_BOX & & By \{DATE\}, the number of peaks at \{HEIGHT\}. &
        \\ \hline
        TS1 & CURRENT & & 10 & CIRCLE & SIZE:10, STROKE\_WIDTH:3, COLOR:\#E84A5F, OPACITY:0.6 & & 
        \\ \hline
        TS1 & CURRENT & & 5 & PAUSE & TIME:10 & & 10 sec.
        \\ \hline
        TS1 & SEARCH & UPTO:28 & 5 & DRAW\_DATA & & & up to 28 days
        \\ \hline
        TS1 & FALL & SLOPE\_LTE:-15 & 5 & TEXT\_BOX &  & By \{DATE\}, the number of cases came down noticeably. We should continue to be vigilant. &
        \\ \hline
        ... & ... & ... & ... & ... & ... & ... & ...
        \\ \hline
    \end{tabular}
\end{table*}

\begin{table*}[t]
    \centering
    \caption{An example feature action table for the ML provenance story as shown in Figure~\ref{fig:MLworkflows}(a).}
    \label{table:ML-feature-action}
    \begin{tabular}{|>{\tiny}m{0.4cm}
                    |>{\tiny}m{1cm} 
                    |>{\tiny}m{1.8cm} 
                    |>{\tiny}m{0.4cm}
                    |>{\tiny}m{1.4cm}
                    |>{\tiny}m{4.1cm}
                    |>{\tiny}m{4.1cm}
                    |>{\tiny}m{1cm}|}
        \hline
        \textbf{ID} &
        \textbf{Feature} & 
        \textbf{Feature Parameters} &
        \textbf{Rank} &
        \textbf{Action} & 
        \textbf{Action Parameters} & 
        \textbf{Text} &
        \textbf{Comment}
        \\ \hline
        TS1 & FIRST  & & 10 & DRAW\_DATA & & &
        \\ \hline
        TS1 & CURRENT & & 10 & TEXT\_BOX & BOX:1 & A newly-trained model achieved testing accuracy of \{TEST\}\% and training accuracy of \{TRAIN\}\%, denoted \{TEST\}\% [\{TRAIN\}\%]. &
        \\ \hline
        TS1 & CURRENT & & 10 & CIRCLE & SIZE:10, OPACITY:0.6, STROKE\_WIDTH:3, COLOR:\#FFA500 &  & 
        \\ \hline
        TS1 & CURRENT & & 10 & NODE & SIZE:5, COLOR:\#FFA500, OPACITY:0.6  &  & 
        \\ \hline
        TS1 & CURRENT & & 10 & TEXT\_POS & X:10, Y:180, COLOR\_TEXT:\#EC5800, COLOR\_BG:\#808080, FONT\_SIZE:13 & Accuracy: \{TEST\}\% &
        \\ \hline
        TS1 & CURRENT  & & 10 & NODE & SIZE:5, COLOR:\#FFA500, OPACITY:0.6 &  & 
        \\ \hline
        TS1 & MAX  & & 10 & DRAW\_DATA & & &
        \\ \hline
        TS1 & CURRENT & & 10 & TEXT\_BOX & BOX:1 & On \{DATE\}, a model achieved the best testing accuracy \{TEST\}\% [\{TRAIN\}\%]. &
        \\ \hline
        TS1 & CURRENT & & 10 & CIRCLE & SIZE:10, COLOR:\#008000, OPACITY:0.6, STROKE\_WIDTH:3, VISIBLE:TRUE & & 
        \\ \hline
        TS1 & CURRENT & & 10 & NODE & SIZE:5, COLOR:\#008000, OPACITY:0.6, VISIBLE:TRUE &  &
        \\ \hline
        TS1 & CURRENT & & 5 & PAUSE & TIME:15 & & 15 sec.
        \\ \hline
        TS1 & LAST  & & 10 & DRAW\_DATA & & &
        \\ \hline
    \end{tabular}
\end{table*}


\section{\textbf{Meta-Storyboard Programs}}
\label{App:Meta-Storyboard}

\subsection{B.1 Generic Meta-Storyboard Program}
Programs for meta-storyboards have very similar structures. A new program can be implemented easily by adapting an existing program. The structure of a meta-storyboard program typically has the following six steps:

\noindent\textbf{STEP 1: Data Selection and Preprocessing}
\begin{itemize}
    \item Based on a user's input, the program selects the data. For example, with COVID-19 data, a user may select a region, the program assigns the time series for the selected region to TS1, which is the ID used in the feature-action table.
    \item For a more complicated storyboard, this step may include processes for creating derived data, e.g., computing the difference between two time series selected by a user.
\end{itemize}

\noindent\textbf{STEP 2: Processing Feature-Action Table}
\begin{itemize}
    \item For each row in the feature-action table, the program 
    \begin{itemize}
        \item detects the specified feature,
        \item registers the specified action against the feature, and
        \item assigns the specified rank to the action.
    \end{itemize}
    \item In addition to the parameters for features and actions, there is a data buffer shared by the feature detection function \texttt{detectFeatures()} and the action registration function \texttt{registerActions()}. The buffer allows \texttt{registerActions()} to access information such as the current data point, the previous data point before the feature detection (excluding the dummy feature CURRENT), and so on.
    \item When there are multiple time series specified in the feature-action table, this step processes all time series concurrently.
\end{itemize}

\noindent\textbf{STEP 3: Integrated Multiple Time Series}
\begin{itemize}
    \item If there are multiple time series, the program creates an integrated time series with combined ranks and all registered actions. The ranks are combined using a Gaussian Mixture Model (GMM).
\end{itemize}
\noindent\textbf{STEP 4: Segmentation and Action Selection}
\begin{itemize}
    \item The program determines the number of time segments according to the length of the animation to be created. The default number is 1 segment, i.e., no segmentation is required.
    \item The program applies a segmentation algorithm to divide the integrated time series according to the specified number of segments.
    \item For each time series, the program selects the important feature-action pairs (with higher ranks) according to the time allowed for the animation of this segment.
\end{itemize}

\noindent\textbf{STEP 5: Create Animated Visualization}
\begin{itemize}
    \item The program displays the selected actions as an animated sequence.
\end{itemize}

\subsection{B.2 Software Prototyping, Adaption, and Generalisation}

We implemented six different meta-storyboards, with different settings in terms of instructions from the meta-authors, application data, target users, feature-action tables, and visual representations. Three meta-authors were involved in creating the three storyboards for COVID-19. One developer wrote programs for prototyping  these meta-storyboards initially, and a second developer ported the programs to the RAMPVIS server with some adaptation. The second developer subsequently implemented three ML storyboards, which were created by two meta-authors. The generic program described in Appendix B.1 was formulated by the developers after bringing all these programs and the associated experience together. We are continuing this development as part of an ML infrastructure. 

\newpage



\end{document}